%% file: ms.tex
\documentclass{emulateapj}
\newcommand{\lapprox }{{\lower0.8ex\hbox{$\buildrel <\over\sim$}}}
\newcommand{\gapprox }{{\lower0.8ex\hbox{$\buildrel >\over\sim$}}}

\slugcomment{DRAFT \today}
\shorttitle{YSOs in B59}
\shortauthors{Covey et al.}

\begin{document}

\title{The Age, Stellar Content and Star Formation Timescale of the B59 Dense Core}

\author{K.~R.~Covey\altaffilmark{1,2,3,4}, C.~J.~Lada\altaffilmark{1}, C.~Roman-Zuniga\altaffilmark{5}, A.~A.~Muench\altaffilmark{1}, J.~Forbrich\altaffilmark{1}, J.~Ascenso\altaffilmark{1}}

\altaffiltext{1}{Harvard-Smithsonian Center for Astrophysics, 60 Garden Street, Cambridge, MA 02138} 
\altaffiltext{2}{Hubble Fellow}
\altaffiltext{3}{Current Address: Cornell University, Department of Astronomy, 226 Space Sciences Building, Ithaca, NY 14853}
\altaffiltext{4}{Visiting Astronomer at the Infrared Telescope Facility, which is operated by the University of Hawaii under Cooperative Agreement no. NCC 5-538 with the National Aeronautics and Space Administration, Science Mission Directorate, Planetary Astronomy Program.}
\altaffiltext{5}{Centro Astron\'omico Hispano Alem\`an/CSIC-IAA, Granada 18006, Spain}

\begin{abstract}  We have investigated the stellar content of Barnard 59 (B59), the most active star-forming core in the Pipe Nebula.  Using the SpeX spectrograph on the NASA Infrared Telescope Facility, we obtained moderate resolution, near-infrared (NIR) spectra for 20 candidate Young Stellar Objects (YSOs) in B59 and a representative sample of NIR and mid-IR bright sources distributed throughout the Pipe.  Measuring luminosity and temperature sensitive features in these spectra, we identified likely background giant stars and measured each star's spectral type, extinction, and NIR continuum excess.  

To measure B59's age, we place its candidate YSOs in the Hertzsprung-Russell (HR) diagram and compare their location to YSOs in several well studied star forming regions, as well as predictions of pre-main sequence evolutionary models.  We find that B59 is composed of late type (K4-M6) low-mass (0.9--0.1 M$_{\odot}$) YSOs whose median stellar age is comparable to, if not slightly older than, that of YSOs within the $\rho$ Oph, Taurus, and Chameleon star forming regions.  Deriving absolute age estimates from pre-main sequence models computed by D'Antona et al., and accounting only for statistical uncertainties, we measure B59's median stellar age to be 2.6$\pm$0.8 Myrs.  Including potential systematic effects increases the error budget for B59's median (DM98) stellar age to 2.6$^{+4.1}_{-2.6}$ Myrs.  We also find that the relative age orderings implied by pre-main sequence evolutionary tracks depend on the range of stellar masses sampled, as model isochrones possess significantly different mass dependences. 

The maximum likelihood median stellar age we measure for B59, and the region's observed gas properties, suggest that the B59 dense core has been stable against global collapse for roughly 6 dynamical timescales, and is actively forming stars with a star formation efficiency per dynamical time of $\sim$6$\%$.  While the $\sim$150\% uncertainties associated with our age measurement propagate directly into these derived star formation timescales, the maximum likelihood values nonetheless agree well with recent star formation simulations that incorporate various forms of support against collapse, such as sub-critical magnetic fields, outflows, and radiative feedback from protostellar heating.
\end{abstract}


\section{Introduction}

The current consensus is that star formation progresses rapidly over a wide 
range of environments and spatial scales.  This view
is supported by observations that a stellar population's 
age spread is rarely larger than twice the system's crossing time 
\citep[$\tau_{cross} = R_{core} / \sigma_v$][]{Elmegreen2000} and that ongoing star 
formation appears ubiquitous within well known molecular clouds: there are few examples 
of clouds in a pre-star forming state, or of clouds associated with stellar 
populations older than $\sim$5 Myrs \citep{Hartmann2001a,Ballesteros-Paredes2007}.  
Taken together, these observations suggest that star formation starts soon
after a molecular cloud forms, and persists only a few 
dynamical times before the cloud is disrupted or dissipates.

The Pipe Nebula \citep{Alves2007} and California Molecular Cloud \citep{Lada2009} 
are anomalies within this prompt star formation paradigm.  Lacking major 
star formation activity, these molecular complexes have only recently begun 
to receive significant scrutiny. The first global study of the Pipe was the CO 
survey by \citet{Onishi1999}, which identified 14 dense clumps that may 
be future sites of active star formation.  \citet{Lombardi2006} then applied 
the NICER technique \citet{Lombardi2001} to 2MASS data to 
construct an extinction map of the Pipe.  This map, which is shown in Figure 1,
leverages the dense screen of background stars provided by the Pipe's projection
against the Galactic bulge, as well as its proximity \citep[d $\sim$130 pc; ][]{Lombardi2006},
to resolve physical structures as small as 0.04pc.  

\begin{figure*}
\epsscale{0.8}
\plotone{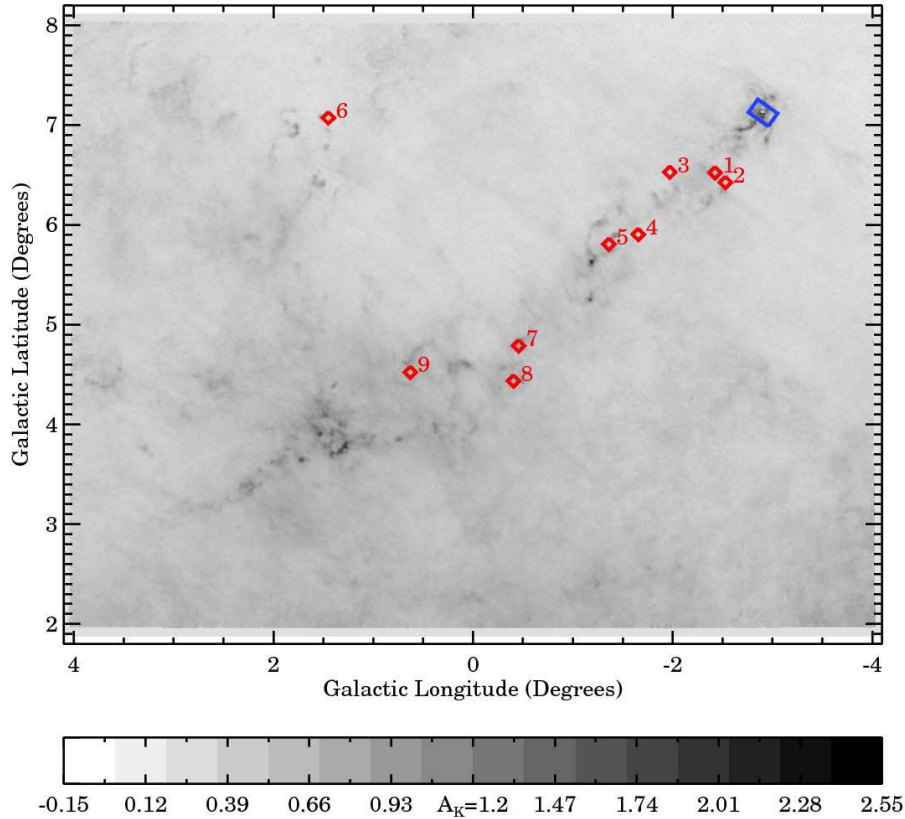}
\caption{ \normalsize{The locations of MIPS bright sources observed in this program (red diamonds), labelled with the [CLR2010] source number from Table \ref{tab:obslog_Pipe_TWHya}. Provided for context is the NICER extinction map of the Pipe Nebula produced by \citet{Lombardi2006} (greyscale contours); the blue box in the upper right identifies the location of B59, shown in more detail in Figure \ref{fig:justb59}.  }}\label{fig:wholepipe}
\end{figure*}

Studies of the Pipe's dense cores have begun to inform our understanding of the 
earliest stages of the star formation process.  Using a wavelet filtering technique 
to identify individual dense cores in the Lombardi et al. extinction map, \citet{Alves2007} 
noted that the Pipe's dense core mass function qualitatively resembles the 
stellar mass functions measured in young clusters \citep[e.g.][]{Muench2002, Luhman2004}
and the field \citep{Covey2008a,Bastian2010}, but with a peak mass that
is larger by a factor of three.  Alves et al. interpreted this as evidence 
that the stellar mass function largely reflects the physics of 
core formation, with a global star formation efficiency of 
$\sim$ 33\% between the dense core and stellar stages.  
Subsequent studies have refined the mass function of Pipe 
cores, adopting physically motivated criteria to identify dense 
cores \citep{Rathborne2009}, and using detailed tests to verify
the fidelity and completeness of the core mass function 
\citep{Kainulainen2009}.  Other authors, however, have emphasized
the difficulty of identifying coherent physical structures from two-dimensional 
column-density maps \citep[e.g.,][]{Smith2008} and cautioned against
making strict one-to-one comparisons between dense core and stellar mass 
functions \citep{Swift2008, Smith2009}.

The physical properties of Pipe dense cores have also shed light on the 
formation and evolution of molecular clouds. A coordinated radio survey 
\citep{Muench2007,Rathborne2008,Lada2008} identified thermal pressure 
as the dominant source of support against collapse for the Pipe's dense cores,
with non-thermal support limited to sub-sonic flows.  With typical sizes of 
0.1 parsec, and measured velocity dispersions of 0.1-0.2 km s$^{-1}$, these 
dense cores possess sound crossing times and minimum ages of $\sim$1 
million years.  These core properties, and particularly their derived lifetimes,
could help discriminate between theoretical models where turbulence 
\citep{Klessen2000, Mac-Low2004, Offner2008} or magnetic fields 
\citep{Nakano1984, Shu1987, Mouschovias1999, Nakamura2008} 
provide the dominant means of support for molecular cores.  This 
evidence, however, is somewhat circumstantial, and relies on a critical 
assumption: that the Pipe's dense core population will one day become
active sites of ongoing star formation.  

The Barnard 59 dense core \citep[hereafter B59; originally dubbed the `Sinkhole' by ][]{Barnard1927} 
could provide a clearer test for these theoretical models: as the only site of active 
star formation within the Pipe, no extrapolations of the core's future behavior are necessary.  
While B59 is the Pipe's most massive core, its physical properties (including 
the dominance of thermal motions) are nonetheless representative 
of the broader population of dense cores within the Pipe, making it 
an instructive case study.  Since 1950, seven optically visible 
H$\alpha$ emitting stars (or unresolved multiple systems) 
have been associated with B59 \citep[V359 Oph, KK Oph AB, LkH$\alpha$ 
345 Oph, LkH$\alpha$ 346 Oph AB, KW 002, TH$\alpha$ 27-5, and LkH$\alpha$ 
347 Oph ;][]{Merrill1950, The1964, Stephenson1977, Kohoutek2003, Herbig2005}.  
A recent Spitzer survey of B59 by \citet{Brooke2007} identified 16 additional
candidate members with strong near-infrared (NIR) excesses, 4 of which were detected previously in the sub-mm 
\citep{Reipurth1996} or by IRAS at far infrared wavelengths.  These 
sources appear to span a range of evolutionary states, from the low-mass, 
relatively embedded Class 0/I protostar studied by \citet{Riaz2009}, to 
more revealed Class II sources.  Analyses of Spitzer, ROSAT, and
XMM observations of the Pipe Nebula confirm that B59 contains the vast majority 
of the Pipe's recently formed stars \citep{Forbrich2009,Forbrich2010}.

The physical properties of the B59 dense core make it a good analog for modern
computational simulations of the collapse of isolated dense cores.  In this respect,
characterizing the timescale and efficiency of star formation in the B59
dense core would provide a good opportunity to compare these empirically 
measured quantities with the predictions of these simulations.  Various methods
have been developed to estimate the ages of pre-main sequence stars, with 
differing advantages and levels of accuracy.  The destruction of a star's primordial
lithium abundance has been used to estimate the ages of pre-main sequence stars with
moderate success \citep[i.e., typical uncertainties of $\sim$10 Myrs for members of 10-30 Myr moving groups; ][]{Mentuch2008}, but this method would be difficult to apply to B59's YSOs: measuring lithium abundances requires high resolution, high S/N optical spectra that would be difficult to acquire for B59's heavily extincted YSOs \citep[A$_V \geq$ 5][]{Brooke2007}, and recent studies indicate that lithium depletion may not a good age indicator for young (t$<$10 Myrs) T Tauri stars \citep{Sestito2008}.  A pre-main sequence star's surface gravity also rises as it contracts towards the main sequence, providing an additional opportunity to constrain the star's age.  Theoretical models suggest that 
a precision of 0.1 dex in log g can provide age estimates accurate to within 50\% \citep[see Figure 8, ][]{Doppmann2005}, but this precision is difficult to achieve in practice.  Log g values have been measured for pre-main sequence stars with an accuracy of $\sim$0.2 dex \citep[e.g., ]{Doppmann2005, Takagi2010}, providing age estimates accurate to $\sim$1-2 Myrs, but as with lithium depletion studies, these measurements depend on precise measurements of line shapes and strengths, requiring observationally intensive high resolution, high signal-to-noise spectra: obtaining spectra of this quality for $\sim$20 B59 YSOs would require multiple nights on an 8-meter class telescope.  Surface gravity indicators have been developed for lower-resolution spectra, but these indicators typically depend on observations at J or shorter wavelengths, again difficult to obtain for B59 YSOs, and have a typical precision of $\pm$0.5 dex, providing somewhat poorer age discrimination than is available with standard HR diagram analyses \citep[$\sim$ 3-5 Myrs][]{Gorlova2003,Allers2009}.

\begin{figure}
\epsscale{1.05}
\plotone{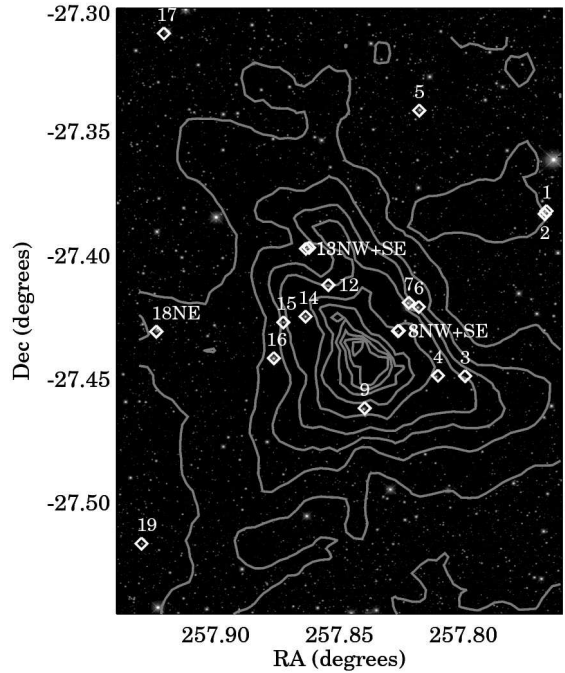}
\caption{ \normalsize{The K band image of B59 obtained by \citet{RomanZuniga2009}, with locations of candidate YSOs observed in this program identified by diamonds and the source number assigned by \citet{Brooke2007}.  The NICER-based extinction map constructed by \citet{RomanZuniga2009} is shown for comparison as white contours.  The lowest contour represents an extinction of A$_V=$2.5; contours correspond to a 5 mag. increase in extinction up to A$_V=$22.5, and increases of 10 mag. thereafter.}}\label{fig:justb59}
\end{figure}

To characterize the timescale, efficiency, and products of star formation 
within the Pipe Nebula, we have conducted the first spectroscopic census of B59's stellar 
population.  Our observations discriminate bona-fide B59 members from background 
giants in the Galactic bulge, and establish that these members are late type (K4-M6) 
low-mass (0.9--0.1 M$_{\odot}$) young stellar objects (YSOs).  We then derive estimates
of the absolute and relative ages of B59's YSOs by comparing their locations within
the observational Hertzsprung--Russell diagram with theoretical models and YSOs in other
well known clusters.  We measure B59's 
median stellar age to be $\sim$2.6$^{+4.1}_{-2.6}$ Myrs; while the error bars on this 
measurement are significant, the maximum likelihood estimate implies 
that star formation has persisted within B59 for more than five times the current
dynamical timescale and crossing time of B59's dense gas.

In Section 2, we present our SpeX observations of candidate B59 
members.  Our method for measuring stellar parameters from these 
spectra is introduced in Section 3, along with our algorithm for 
placing those objects in the HR diagram and inferring ages and 
masses from pre-main sequence evolutionary models.  We discuss our 
results in Section 4, and summarize our findings in Section 5. We also 
include a NIR spectral atlas of TW Hydrae members as an appendix to this work.

\input{t1}

\section{Observations}\label{sec:Obs}

Using the SpeX spectrograph \citep{Rayner2003} at NASA's 2.5 meter 
Infrared Telescope Facility, we obtained low (R$\sim$200) 
and moderate (R $\sim$ 2000) resolution spectra of candidate 
young stellar objects (YSOs) in B59 and the Pipe Nebula.  Observations
were carried out during the early mornings of April 14-18 2008 (HST); typical 
seeing was 0.7\arcsec, with clear skies interrupted by periods of light cirrus.  
The bulk of our observations were carried out in 
SpeX's SXD mode with the 0.3\arcsec slit, providing nearly contiguous 
spectral coverage from 0.8 to 2.5 microns with a 
resolution of R$\sim$2000.  During a night of abnormally
poor seeing ($\sim$1\arcsec) we made use of the larger 0.5\arcsec slit, degrading the
spectral resolution to R$\sim$1200.  

For maximum observational efficiency, we chose not use the image rotator 
to align the slit with the 
parallactic angle over the course of each night.  Spectra 
of telluric standards taken at slit rotations that differ by 30-60$^{\circ}$ 
suggest that color-dependent slit losses due to atmospheric refraction 
are relatively minor, altering the slope of the spectral energy 
distribution (SED) between the J and K bands by less than 10\%.  During our last night of observations we
also utilized SpeX's high-throughput, low-resolution (R$\sim$200) prism mode
to obtain high fidelity NIR SEDs for candidate YSOs in B59.  During these
observations, care was taken to align the slit to within 10$^{\circ}$ of
the parallactic angle, except in cases where the slit was rotated to allow
observations of individual components of an apparent multiple system.  

\citet{Brooke2007} identified candidate YSOs in B59 
by searching for IR excesses in spectral energy distributions 
(SEDs) constructed from 2MASS, IRAC and MIPS 
photometry. For the remainder of this paper, we identify these 
candidate YSOs with a [BHB2007] prefix, followed by 
the source ID assigned by Brooke et al. in their Table 1.  
We obtained spectra for each of these candidate
B59 members, with the exception of two objects 
too faint to observe with SpeX ([BHB2007] 10, \& 11).  A K 
band image of B59 observed by \citet{RomanZuniga2009} 
is shown in Figure \ref{fig:justb59}, overplotted with the locations
of the candidate YSOs observed in this program. 
Our SXD observations are summarized in Table \ref{tab:obslog_B59},
where we indicate the slit width and exposure time
used to obtain each source's spectrum.  We include for completeness
the 2MASS position and magnitudes for each source as well as H and K 
band magnitudes measured by \citet{RomanZuniga2009} from their NTT 
images of the cluster.  We also include positions and 2MASS magnitudes for 3 likely 
B59 members (V359 Oph, KK Oph AB, and LkH$\alpha$ 345 Oph) 
with spectral types in the literature that suffice to place
each star in the HR diagram.

To characterize the more extended population of bright IR objects that extends across the Pipe, we obtained spectra for a sample of objects with K$<$10 (to ensure accessibility to SpeX) and detections at both 24 and 70 microns in an extensive Spitzer survey of the Pipe Nebula \citep{Forbrich2009}.  The locations of these sources are indicated in Figure 
\ref{fig:wholepipe}, and their NIR properties are listed in the top tier of Table \ref{tab:obslog_Pipe_TWHya}.  
 We observed all four sources ([CLR2010]\footnote{For consistency with the [BHB2007] designation adopted for the candidate B59 members identified by \citet{Brooke2007}, we adopt a source designation of [CLR2010], followed by the source number given in Table \ref{tab:obslog_Pipe_TWHya}, for the IR bright sources we observed throughout the Pipe.} 1, 2, 5 and 7) that met our flux criteria and are located within high extinction peaks identified by \citet{Alves2007}, as we considered these sources the most likely bona fide YSO candidates. We additionally observed 6 similar sources that are not associated with extinction cores ([CLR2010] 3, 4, 6, 8 and 9).  
 According to the criteria of \citet{Forbrich2009}, two of the sources in our sample ([CLR2010] 1 \& 2) are MIPS excess sources. A third source, [CLR2010] 7, has a MIPS spectral index marginally qualifying as class II but is not listed by \citet{Forbrich2009} due to a position mismatch at the source detection stage.

\input{t2}

The SpeX spectral atlas\footnote{available online at\\ \url{http://irtfweb.ifa.hawaii.edu/$\sim$spex/spexlibrary/IRTFlibrary.html}} \citep[][Rayner et al., in prep; Vacca et al., in prep.]{Cushing2005} samples many spectral types and 
luminosity classes.  To supplement this grid of standards with
stars possessing pre-main sequence surface gravities, we observed 
members of the TW Hydrae moving group and $\rho$ Ophiuchus cluster. 
The pre-main sequence standards are listed in the bottom tier of Table 
\ref{tab:obslog_Pipe_TWHya}, and spectra of the TW Hya members are shown 
in Appendix \ref{app:TWHyaatlas}.

All data were reduced using SpeX's dedicated IDL reduction package, SpeXtool 
\citep{Cushing2004}.  Telluric absorption was removed using
observations of A0V stars at airmasses similar  
to those of the target ($\Delta$sec$ z \lesssim$0.1).  Hydrogen features 
were removed from the A0V spectra by dividing by a model of Vega, and
the target's true spectral slope and absolute flux densities were recovered by 
multiplying the target spectrum by a $\sim$10,000 K blackbody spectrum 
scaled to match that of the telluric standard \citep{Vacca2003}. Tables \ref{tab:obslog_B59}
 and \ref{tab:obslog_Pipe_TWHya} list the signal-to-noise (S/N) levels
 measured in the J, H and K$_s$ bands of each source's reduced SXD 
spectrum.  Two S/N estimates are listed for each band: the first 
estimate is the S/N calculated directly by SpeXtool.  The 
second estimate is an empirical lower limit to each spectrum's S/N
generated from the standard deviation of relatively featureless regions in each band 
(J: 1.215-1.235 $\mu$m, H: 1.52-1.54 $\mu$m, K: 2.13-2.15 $\mu$m).  

\section{Analysis}

\input{t3}

We have analyzed these SpeX spectra to distinguish bona fide B59 
members from contaminating background giants, and to infer the 
physical properties of each YSO.  
In \S \ref{estAvandR}, \ref{measurespt} and \ref{derived_types}, we 
describe an iterative technique that identifies the spectral type, extinction 
and veiling that best reproduce the individual features and overall shape 
of each candidate B59 YSO's spectrum.  In \S \ref{estimatelum}, 
we supplement these spectral measurements with photometry from the literature 
to estimate the effective temperature (T$_{eff}$) and bolometric luminosity 
(L$_{bol}$) of each target.  Placing these YSOs in an observational 
Hertzsprung-Russell (H-R) 
diagram, we used pre-main sequence evolutionary models to infer
masses and ages for each object.  We present in \S \ref{deriveage} an 
estimate of B59's absolute age, as well as a relative age
estimate derived from comparison to other clusters
with similarly homogeneous photometric and spectroscopic datasets.

\subsection{Estimating Extinction \& Veiling From Synthetic Photometry}
\label{estAvandR}

Following \citet{Meyer1997}, we assume that a YSO's observed NIR color (e.g., $(J-H)_{obs}$) 
is due solely to its intrinsic photospheric color (e.g., 
$(J-H)_{int}$), an additional continuum excess from its circumstellar 
disk (e.g., $\Delta(J-H)$, such that $(J-H)_{CTTS} = (J-H)_{int}$ + $\Delta(J-H)$), 
and an independent extinction/reddening term (e.g., $E(J-H)$ = 0.613 $\times$ A$_H$; 
$(J-H)_{obs}$ = $(J-H)_{CTTS} + E(J-H)$).  Algebraically combining 
the \citet{Meyer1997} CTTS locus (transformed onto the 2MASS photometric 
system) with a well behaved extinction law\footnote{\citet{Roman-Zuniga2007} measured the near/mid-IR extinction law 
in B59 itself, identifying A$_J$/A$_K$ and A$_H$/A$_K$ ratios that 
agree within 1$\sigma$ of the values measured by \citet{Indebetouw2005}
from large numbers of 2MASS stars. Given the agreement between these two
studies, we adopt the \citet{Indebetouw2005} extinction law for maximum
compatibility with the 2MASS photometry we utilize to characterize YSOs
in B59 and other star formation regions.} then requires
a unique combination of reddening and disk excesses/veiling to
reconcile a YSO's intrinsic and observed \textit{(J-H)} and \textit{(H-K$_s$)} colors:

\begin{eqnarray}
A_H = 2.512 \times (J-H)_{obs} - \nonumber \\
1.527 \times (H-K_s)_{obs} - 1.254 \\ \nonumber \\
\Delta(J-H) = (J-H)_{obs} - \nonumber \\
0.613 \times A_H - (J-H)_{int} \\ \nonumber \\
\Delta(H-K_s) = (H-K_s)_{obs} - \nonumber \\
0.354 \times A_H  - (H-K_s)_{int}
\end{eqnarray}

YSOs often demonstrate considerable photometric variability, including in the NIR regime we focus on here \citep[e.g., ][]{Carpenter2001, Morales-Calderon2009}. To estimate each YSO's extinction and NIR excess
at the exact epoch that the spectra were obtained, 
we calculate synthetic NIR photometry from each target's 
SpeX spectrum following the prescription presented by 
\citet{Covey2007} in their \S 3.3.1.  Using these JHK 
magnitudes, and assuming intrinsic photospheric colors
characteristic of K7/M0 type stars, we can
therefore obtain a good initial estimate of each
YSO's extinction (A$_H$) and $\Delta$(J-H) and $\Delta$(H-K$_s$) 
color excesses.  

\subsection{Extinction and Veiling Resistant Spectral Classification}\label{measurespt}

Numerous investigators have identified H and K band spectral indices
sensitive to a star's luminosity class and spectral type 
\citep{Kleinmann1986,Ali1995,Wallace1997,Meyer1998,Aspin2003,Ivanov2004,Allers2007}.  
We present in Table \ref{tab:lineindices} a set of indices well suited for measuring stellar spectral types from 
moderate resolution (R$\sim$1000--2000) spectra.  
Each feature's local continuum is measured from a linear fit to the 
median flux in two neaby regions (`Cont. 1' and `Cont. 2' in 
Table \ref{tab:lineindices}).  Integrating the difference between the expected continuum and the actual flux in the line (defined by the `Line Center and `Line Width' columns of Table \ref{tab:lineindices}), and dividing by the continuum 
flux density in the center of the line region, provides an Equivalent
Width (EqW) measure of the strength of each feature.  Equivalent width measures are 
convenient for spectral classification of YSOs: equivalent widths are relatively insensitive 
to extinction, which affects both the line and the local
continuum equally, and ratios of closely separated lines are relatively robust to 
excess continuum flux, as both lines will be diluted
equally and the sum effect will cancel out.  

The strength and shape of water absorption in the H and K bands
has also been identified by numerous authors as a useful spectral type and luminosity
class indicator \citep[e.g., ][]{Wilking1999,Reid2001,Geballe2002,McLean2003,Slesnick2004,Allers2007,Weights2009}, but is less amenable to characterization as an 
Equivalent Width.  We defined two indices to characterize 
water absorption in the H and K bands; these indices are calculated as:

\begin{equation}
\rm{Index} = \frac{<Band 1>/<Band 2>}{<Band 2>/<Band 3>}
\end{equation}
\noindent where $<$X$>$ simply denotes the mean flux level within the
 range of each band as given in Table \ref{tab:waterindices}.

To account for any residual sensitivity to extinction or veiling
in our spectral indices, we compared our target's index values 
to those measured from artificially reddened and 
veiled spectral standards.  As described in \S \ref{estAvandR},
synthetic photometry from each SpeX spectrum produces an initial estimate 
of each target's extinction and (J-H)/(H-K) color excesses assuming the standard K7/M0 spectral type.   
Adopting the characteristic J band veiling (r$_J$=$\frac{F_{ex}}{F_*} = 0.25$) 
measured by \citet{Cieza2005} for a large sample of CTTSs sets
the initial absolute scale for the veiling fluxes implied by each
source's $\Delta$(J-H) and $\Delta$(H-K$_s$) color excesses:

\begin{eqnarray}
\Delta J = 2.5 \times Log(r_J + 1) \nonumber \\
\Delta H = \Delta J + \Delta(J-H) \nonumber \\
r_H = 10^{\frac{\Delta H}{2.5}}-1
\end{eqnarray}

\begin{figure*}
\epsscale{1.1}
\plotone{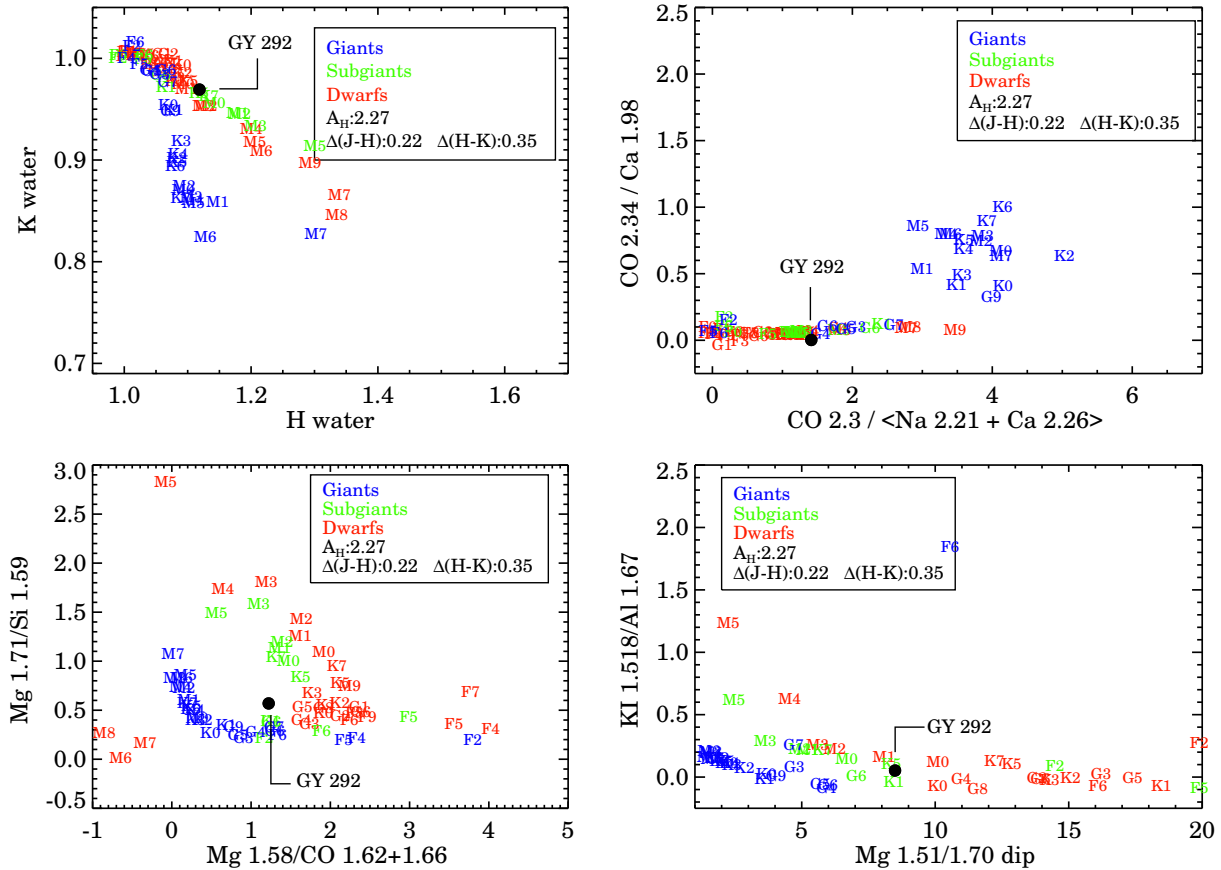}
\caption{ \normalsize{Two dimensional spectral classification
of GY 292: each panel shows luminosity and gravity
sensitive H and K band spectral indices presented in \S \ref{measurespt}.
Indices measured from GY 292, a confirmed YSO with a previously measured 
K8 spectral type \citep{Luhman1999}, are shown as a filled black 
point, while indices measured from spectral standards are
labeled according spectral type, and color-coded
according to luminosity class.  Standard spectra were artificially veiled and extincted to match estimates from GY 292's photometry and spectra before indices were measured.  GY 292 appears to be an $\sim$K5 object with subgiant-like surface gravity, as expected
for a bona-fide YSO. }}\label{fig:sptindices}
\end{figure*}

\begin{figure*}
\epsscale{1.1}
\plotone{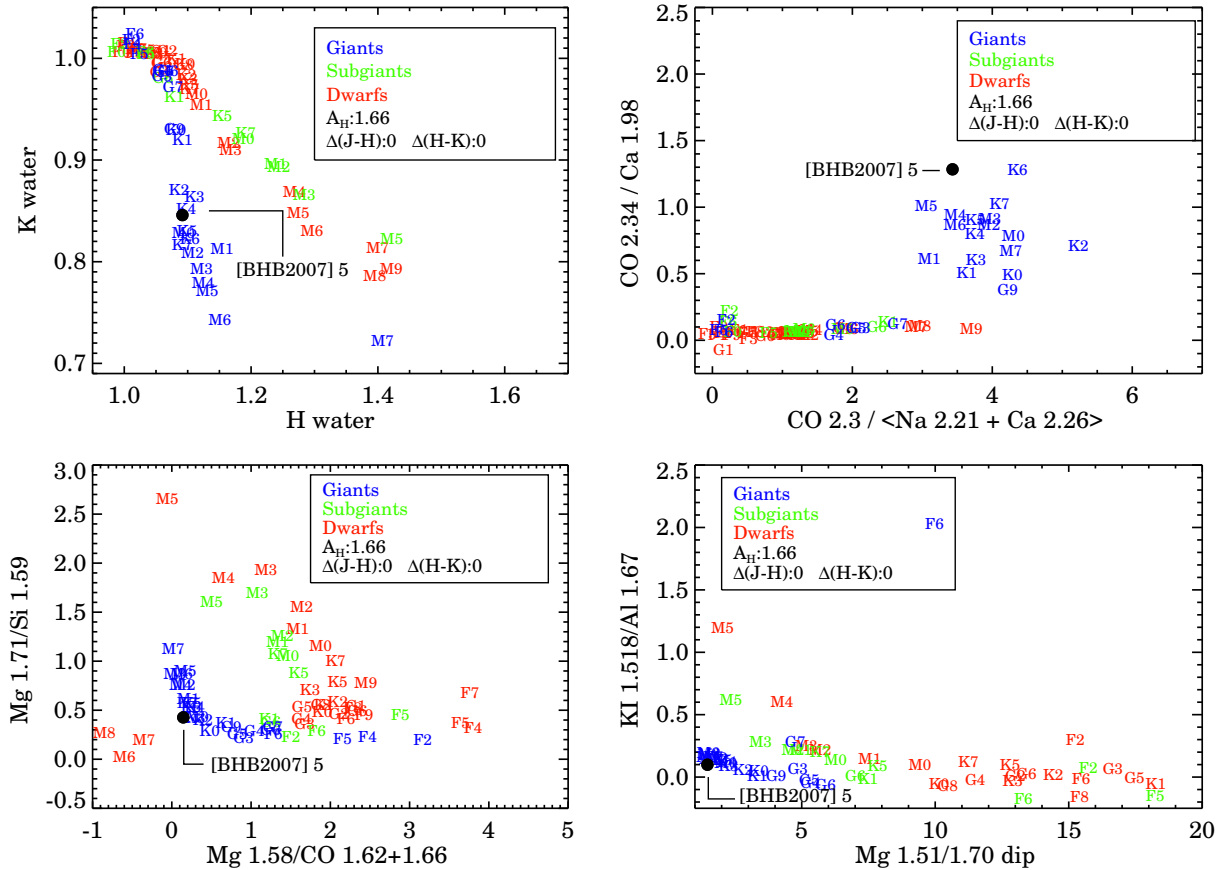}
\caption{ \normalsize{The two dimensional spectral classification
plots for candidate YSO [BHB2007] 5; symbols are as in Figure 
\ref{fig:sptindices}.  [BHB2007] 5 appears to be
a mid-late K star, with giant-like surface gravity; visual inspection
of its spectrum confirms it as a background giant rather than a bona-fide 
YSO (see Fig. \ref{fig:giantspecs}). }}\label{fig:sptindices_B5}
\end{figure*}

We then iteratively adjusted each target's assumed absolute J band veiling 
and spectral type (and thus intrinsic photospheric colors)
to identify the self-consistent spectral type/extinction/veiling combination 
that best replicated the target's measured spectral indices and overall SED shape.
Figures \ref{fig:sptindices} and \ref{fig:sptindices_B5} demonstrate our
two dimensional spectral classification of GY 292 and [BHB2007] 5, 
respectively.    GY 292 appears to be moderately veiled ($\Delta$(J-H): 0.22;
$\Delta$(H-K$_s$): 0.35), but [BHB2007] 5 shows no evidence for veiling, 
so these two stars demonstrate the sensitivity of these spectral 
indices to the presence of veiling.  Veiling has the largest impact on the two water indices shown in the upper left panels of Figures \ref{fig:sptindices} and \ref{fig:sptindices_B5}: standards
veiled to match GY 292 are significantly compressed compared to the
unveiled standards used to type [BHB2007] 5.   

\input{t4}

Figure \ref{fig:sptindices_B5} also demonstrates the utility
of two-dimensional spectral classification for identifying
background giants, as [BHB2007] 5 is clearly identified in each plot
as a giant star; lacking significant veiling, or emission
lines indicative of mass accretion, this object appears to be a
background late K/early M giant star. 

\subsection{Derived Spectral Types}
\label{derived_types}

\begin{figure}
\epsscale{1.2}
\plotone{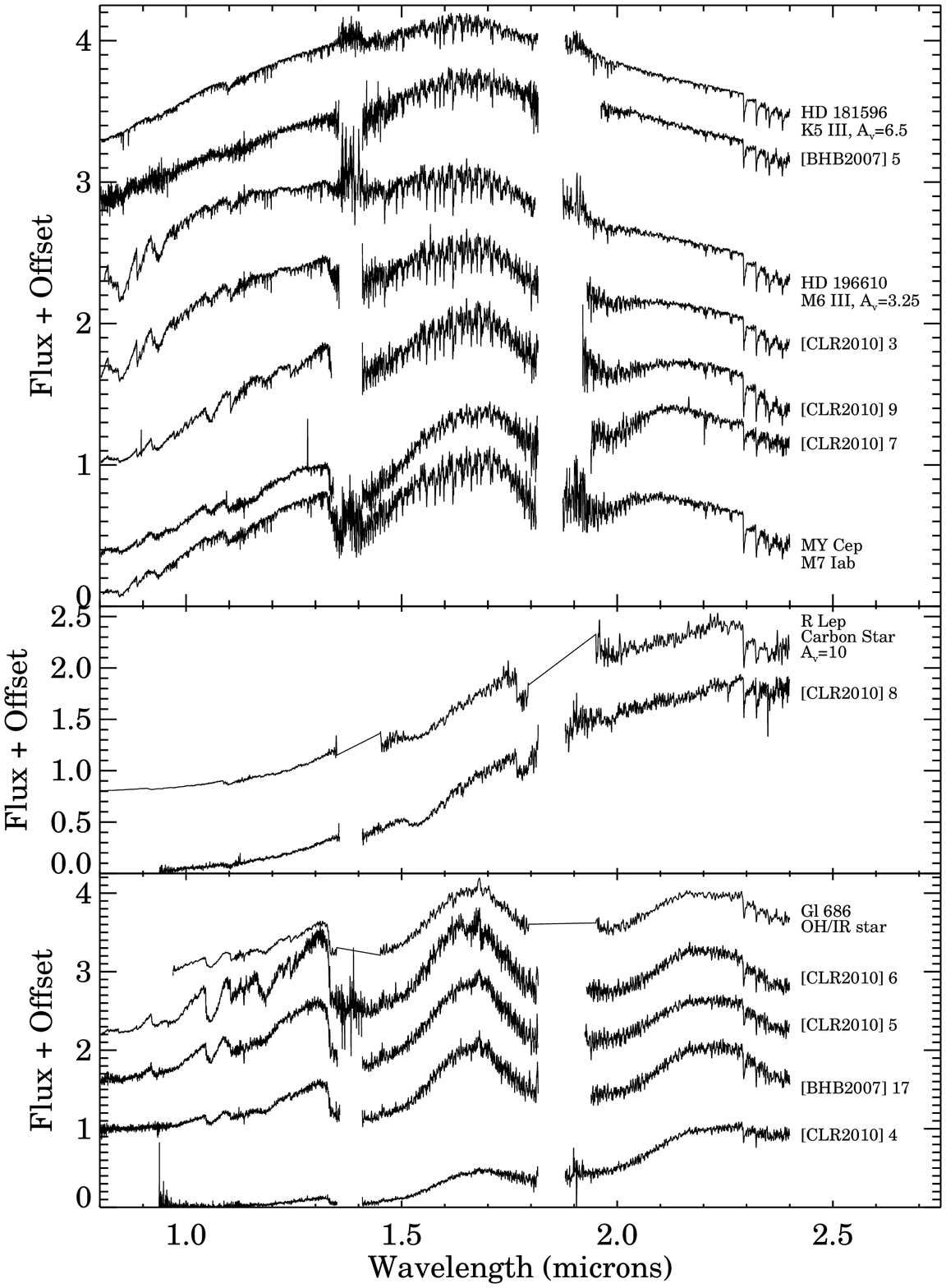}
\caption{ \normalsize{Spectroscopically identified giant stars in B59 and the Pipe Nebula. \textit{Top:} K \& M giants and supergiants -- [BHB2007] 5, [CLR2010] 3, 7, \& 9, shown in comparison to spectra of HD 181596 (K5 III), HD 156014 (M6 III), and MY Cep (M7 I) from the IRTF spectral library \citep{Cushing2005}. HD 181596 and HD 196610 have been artificially reddened to better match the target spectra.  \textit{Middle:} Carbon stars -- [CLR2010] 8 shown for comparision with the reddened spectrum of R Lep \citep{Lancon2000}. \textit{Bottom:} OH/IR stars -- [CLR2010] 4, 5 \& 6, as well as [BHB2007] 17, shown for comparison with the spectrum of OH/IR star Gl 686, from \citet{Lancon2000}. }}\label{fig:giantspecs}
\end{figure}

Applying our classification algorithm to these spectra identifies two 
of the B59 candidate YSOs and the majority of the IR excess stars 
throughout the Pipe as likely giants: we classify
[BHB2007] 5, [CLR2010] 3 and 9 as reddened giants or supergiants, 
[CLR2010] 8 as a giant carbon star, and [CLR2010] 4, 5, 6 and [BHB2007] 17 
as OH/IR stars, likely residing in the Galactic Bulge. 
The spectra of these stars are shown in Figure \ref{fig:giantspecs} along 
with spectra of known giants obtained by \citet{Lancon2000} and 
\citet{Cushing2005}.  Our classification of [BHB2007] 5 and 17 as background
giants is reinforced by their non-detection at X-ray wavelengths in an XMM pointing
containing B59 (Forbrich et al. 2010, in press).

\begin{figure*}
\epsscale{1.0}
\plotone{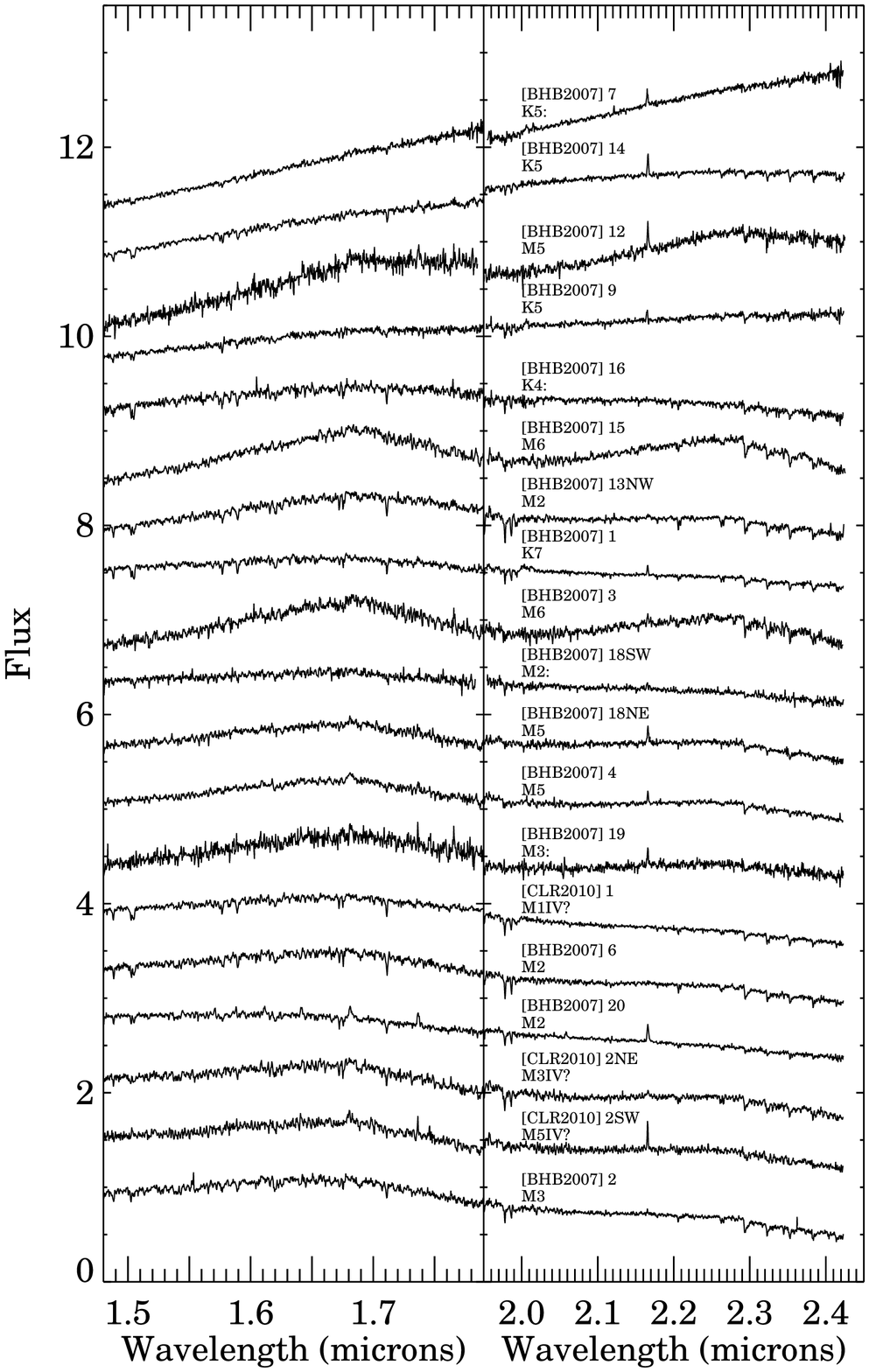}
\caption{ \normalsize{H and K band spectra of confirmed or candidate B59 members.  [BHB2007] 8NW \& SE are not shown, as their spectra are very low signal-to-noise.}}\label{fig:tarspecs}
\end{figure*}

\begin{figure}
\epsscale{1.2}
\plotone{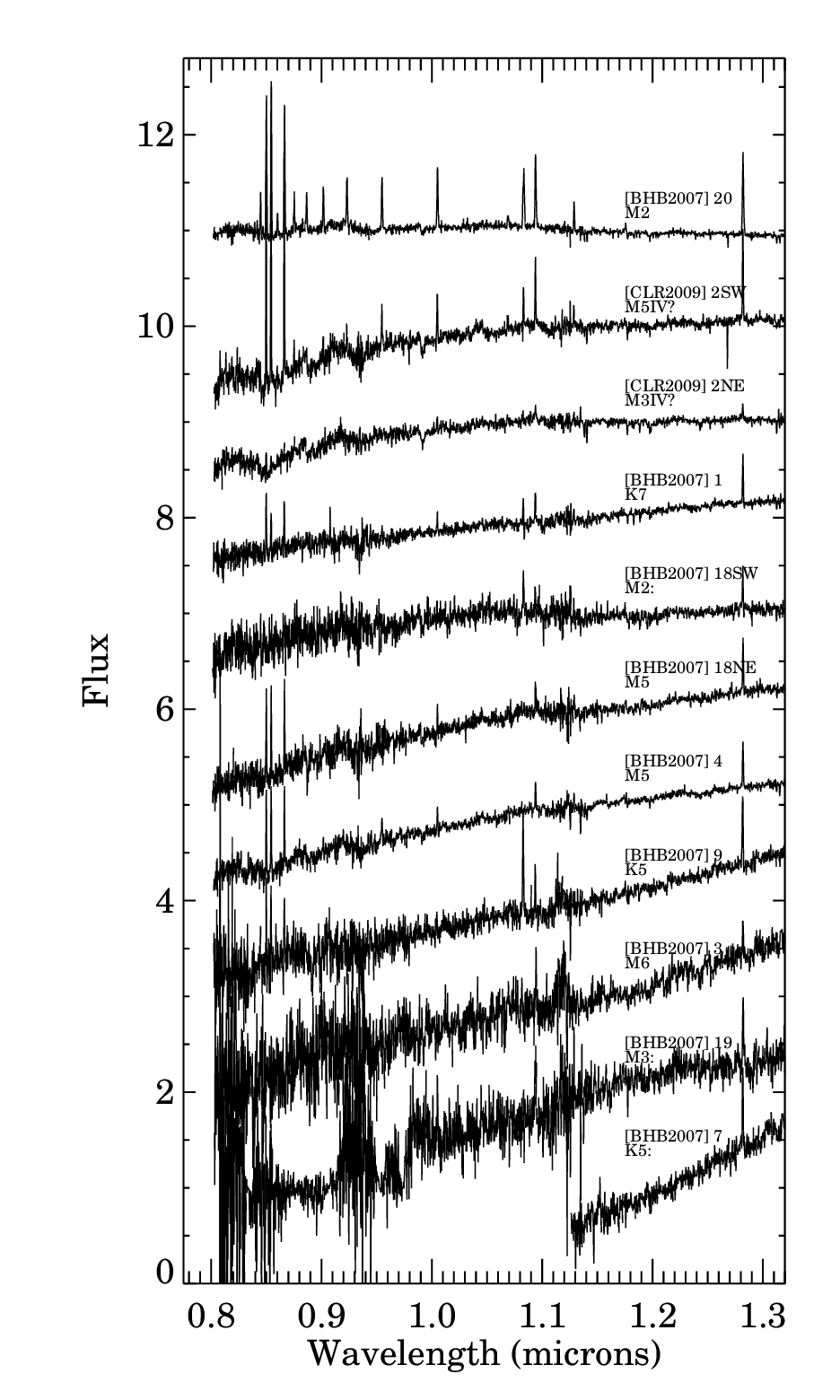}
\caption{ \normalsize{J band spectra of confirmed or candidate B59 members with visible J band emission lines.}}\label{fig:tarspecs-J}
\end{figure}

The spectra of the confirmed B59 YSOs are shown in Figures 
\ref{fig:tarspecs} and \ref{fig:tarspecs-J}, with their derived spectral types, extinctions, 
and JHK veilings presented in Table 
\ref{tab:B59_spts}.  We tested the accuracy of these results
by comparing the parameters we measure for four $\rho$ Oph YSOs to those
reported by \citet{Luhman1999}.  
We derived spectral types for GY 292 (K5), GY 17 (K5), GY 250 (M1) 
and GY 93 (M5); \citet{Luhman1999} derived spectral types
for these objects of K8, K6, M0, and M4 respectively.  
The types assigned to GY 17, 93, and 250 agree within 1 subclass, 
while the type we assign to GY 292 is earlier by 
two/three subclasses \citep[following][our classification scheme omits types K6 and K8]{Kirkpatrick1991}.  
On the basis of this comparison, we adopt a
conservative uncertainty of $\pm$2 subclasses in our derived spectral types.
Similarly, comparing the extinctions and veilings we derive with those
reported by \citet{Luhman1999} suggest that we achieve $\sigma_{A_H} \approx$ 0.5 mag and $\sigma_{r_K} \approx$ 0.35.  

\input{t5}

Two sources, [CLR2010] 1 and 7, present somewhat ambiguous 
spectra.  [CLR2010] 7 possesses a sub-giant/giant-like surface 
gravity, with prominent Paschen $\beta$, Paschen $\gamma$, 
and Brackett $\gamma$ emission features as seen in some YSOs.  
[CLR2010] 7 is also located on the edge of a Pipe extinction 
core, and is on the lower boundary defined by \citet{Forbrich2009} to select candidate class II sources based on their MIPS spectral index.  While these properties are consistent
with a YSO classification, \citet{Ojha2007} identify [CLR2010] 7 as a likely high mass-loss 
AGB star in the Galactic Bulge in their analysis of detections by the Mid-course Space Experiment.  
This classification is also consistent with the observed properties of [CLR2010] 7, explaining its strong 
HI emission as a signature of strong stellar winds.  We tentatively classify this source as a giant star, and include
its spectrum in Figure \ref{fig:giantspecs}.  [CLR2010] 1 is more ambiguous: this source possesses a low (sub-giant) surface gravity, and has a location in the HR diagram consistent with that of other B59 members (see \S 3.3 for a discussion of HR diagram placement).  This HR diagram location implies a rather young age for [CLR2010] 1, however, which is somewhat inconsistent with the source's lack of NIR excess/disk emission or mass accretion sensitive emission lines.  Indeed, [CLR2010] 1's NIR photometry is also consistent with that of a reddened background giant/sub-giant.  Definitively resolving [CLR2010] 1's status will require additional observations; for now we include it in Table \ref{tab:B59_spts} and Fig. \ref{fig:tarspecs} as a potential Pipe YSO, but exclude it from our population analysis of B59 members.

One of the IR excess stars located outside of B59, however, is likely 
a newly identified YSO. [CLR2010] 2 lies in the stem of the Pipe,
within an \citet{Alves2007} extinction core (see Fig. \ref{fig:wholepipe} for its exact location), and was identified by \citet{Forbrich2009} as a new candidate YSO (source 16 in their object list).  While detected as a single source by MIPS, [CLR2010] 2 proved to be a visual binary when observed with SpeX: we
obtained spectra of both the NE and SW components, and the spectral
indices measured from both sources are consistent with pre-main
sequence surface gravities.  Strong HI, HeI, and Ca II emission lines
are also visible in the spectrum of [CLR2010] 2 SW, consistent 
with the source being an actively accreting YSO.  This source
was also identified as an H$\alpha$ emission line source
by \citet{Kohoutek2003}, indicating that the HI emission 
persists over long timescales.  These observations suggest
that [CLR2010] 2 SW is a bona fide YSO associated with the Pipe
Nebula, and [CLR2010] 2 NE is a promising candidate as well.  
The detection of H$\alpha$ from [CLR2010] 2, and the low extinction 
estimates we derive from our SpeX observations, indicate that 
follow-up optical spectroscopy of the 6707$\AA$ Li line may be 
a tractable means of definitively resolving each component's pre-main sequence 
status.  As with [CLR2010] 1, we include [CLR2010] 2 NE and SW in Table \ref{tab:B59_spts} and Figs. \ref{fig:tarspecs} and \ref{fig:tarspecs-J}.

\subsection{Calculating T$_{eff}$ and L$_{bol}$}\label{estimatelum}

To compare the properties of B59's YSOs with theoretical models of pre-main sequence 
evolution, we adopted the SpT to T$_{eff}$ conversion developed by \citet{Luhman1998}.  The T$_{eff}$ implied by each source's
spectral type is shown in Table \ref{tab:inferred_params}.  In 
the spectral type range occupied by most B59 members, our $\pm$2 
subclass spectral type uncertainty corresponds to an $\sim300$ 
K uncertainty in temperature.

With extinction and veiling estimates for each candidate 
B59 member, we are able to calculate stellar luminosities
from their NIR photometry.  We followed \citet{Hillenbrand1997}
in calculating each YSO's bolometric luminosity as: 

\begin{eqnarray}
Log(L_{tot}/L_{\odot}) = 0.4 \times \nonumber \\
\Big(4.75- \big(H - A_H - DM + BC_H \big) \Big) 
\end{eqnarray}

\noindent where DM is the distance modulous to B59 \citep[d=130 pc, 
DM = 5.569; ][]{Lombardi2006} and BC$_H$ is the H band
bolometric correction for each star's spectral type as tabulated 
by \citet{Kenyon1995}; the values adopted for
each star are included for reference in Table \ref{tab:inferred_params}.  

We also calculate the purely stellar component 
of each YSOs bolometric luminosity by removing the 
H band veiling flux, expressed in magnitudes (see \S \ref{measurespt} for the transformation between the veiling flux ratio in a given band (e.g, r$_H$) and the photometric excess in that band (e.g., $\Delta$ H): 

\begin{eqnarray}
Log(L_{stel}/L_{\odot}) = 0.4 \times \nonumber \\
\Big(4.75- \big(H + \Delta H - A_H - DM + BC_H \big) \Big) 
\end{eqnarray}

We adopt the measured 2MASS H band 
magnitude for isolated B59 members in this calculation, 
enabling us to perform an identical 
analysis of YSOs with known spectral types in other well 
studied star forming regions.  Objects with nearby ($<$4\arcsec) 
companions may be blended in 2MASS due to that survey's relatively 
large pixels ($\sim$2\arcsec); for those sources ([BHB2007] 2, 8 NW 
\& SE, 13 NW, 18 NE \& SW; [CLR2010] 2 NE \& SW), we adopt 
the synthetic H band magnitude measured from our spectra, noting that
the SpeXTool reduction pipeline produces spectra with a first-order 
correction for slit losses derived from the observations of the telluric standard.

\begin{figure*}
\epsscale{0.8}
\plotone{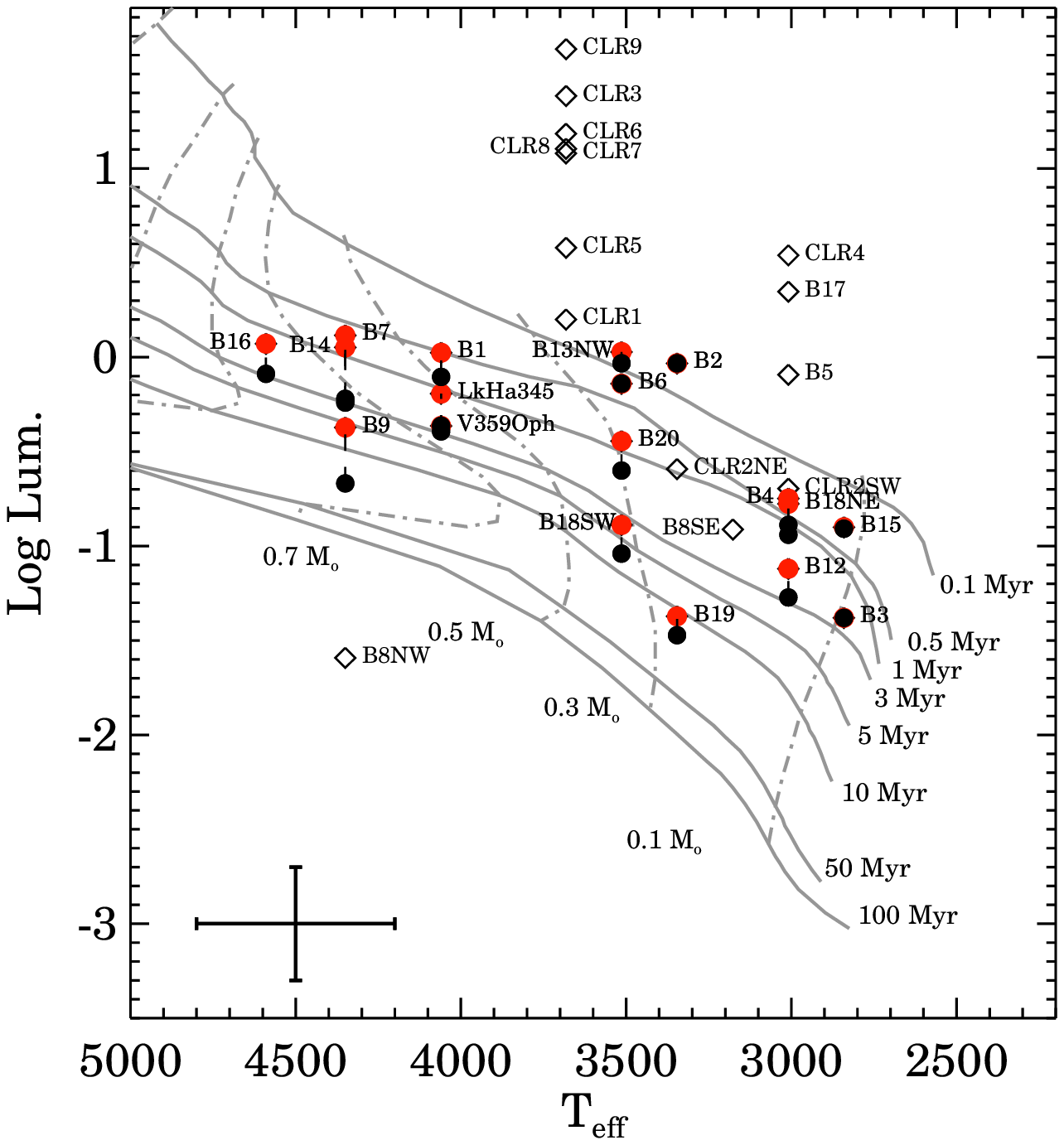}
\caption{ \normalsize{B59 HR diagram calculated using JHK photometry.  Likely B59 members (including previously studied members V359 Oph \& Lk H$\alpha$ 345) are shown twice; red dots indicate their total bolometric luminosity, and black dots indicate their stellar luminosity after subtracting H band veiling.  Objects classified as background giants are shown with open diamonds, as are [BHB2009] 8SE \& NW, whose spectra are too noisy for reliable HR diagram placement.  The DM98 model grid is shown for comparison as solid and dotted lines.}}\label{fig:B59HR}
\end{figure*}

Both luminosity estimates (total and stellar-only) are listed in Table 
\ref{tab:inferred_params} and shown in Figure \ref{fig:B59HR}.
Accounting for the uncertainties in each of the input parameters 
results in a total uncertainty for these luminosity estimates of 
0.3 dex.  The difference between the total and stellar luminosity 
estimated for a given YSO is usually less than 0.2 dex, indicating 
that the veiling correction, while not negligble, does not exceed 
the overall error in each YSO's location in the HR diagram.

\input{t6}

\subsection{Inferring B59's Median Stellar Age}\label{deriveage}

We have used the pre-main sequence (PMS) evolutionary models calculated by 
\citet{Dantona1994} and updated in 1998 (DM98 hereafter), and 
those calculated by \citet{Baraffe1998}, to infer the 
age and mass of each candidate B59 YSO.  These parameters 
were determined by interpolating between model 
points with temperatures and luminosities similar to that
of each YSO.
These theoretical models do not account for accretion luminosity, 
so for this comparison we adopted the veiling subtracted value of L$_{stel}$ for each YSO.
Table \ref{tab:inferred_params} presents masses and ages 
derived for candidate B59 members with reliable spectroscopic 
and photometric information.  

The stellar ages presented in Table \ref{tab:inferred_params} allow
us to identify the characteristic timescale over which B59 has 
been actively forming stars.  We calculate this timescale as 
the median age of all likely B59 members with 
reliable age estimates (i.e., all YSOs listed in Table 
\ref{tab:inferred_params} except [CLR2010] 2 NE \& SW, which 
may not be linked to B59).  We prefer the median to a standard mean 
because it is more robust to outliers with extreme ages (e.g., [BHB2007] 19).  
B59's median stellar age, as inferred from DM98 PMS tracks using
veiling-corrected 
luminosity estimates ($L_{stel}$), is 2.6 Myrs.

Our confidence in the age estimates we calculate here
is reinforced by our ability to independently reproduce the
age estimate derived by \citet{Luhman1999} for their sample of $\rho$ Oph members.  Following \citet{Luhman1999} by comparing 
each star's total (non-veiling corrected) luminosity to the 
DM98 tracks, and adopting their estimated distance of 166 
pc. ($m-M=6.1$), we measure a median stellar age for $\rho$ Oph of 0.21 Myrs, 
consistent with the $\sim$0.3 Myr median stellar age  
\citet{Luhman1999} derived.

\begin{figure*}
\epsscale{0.8}
\plotone{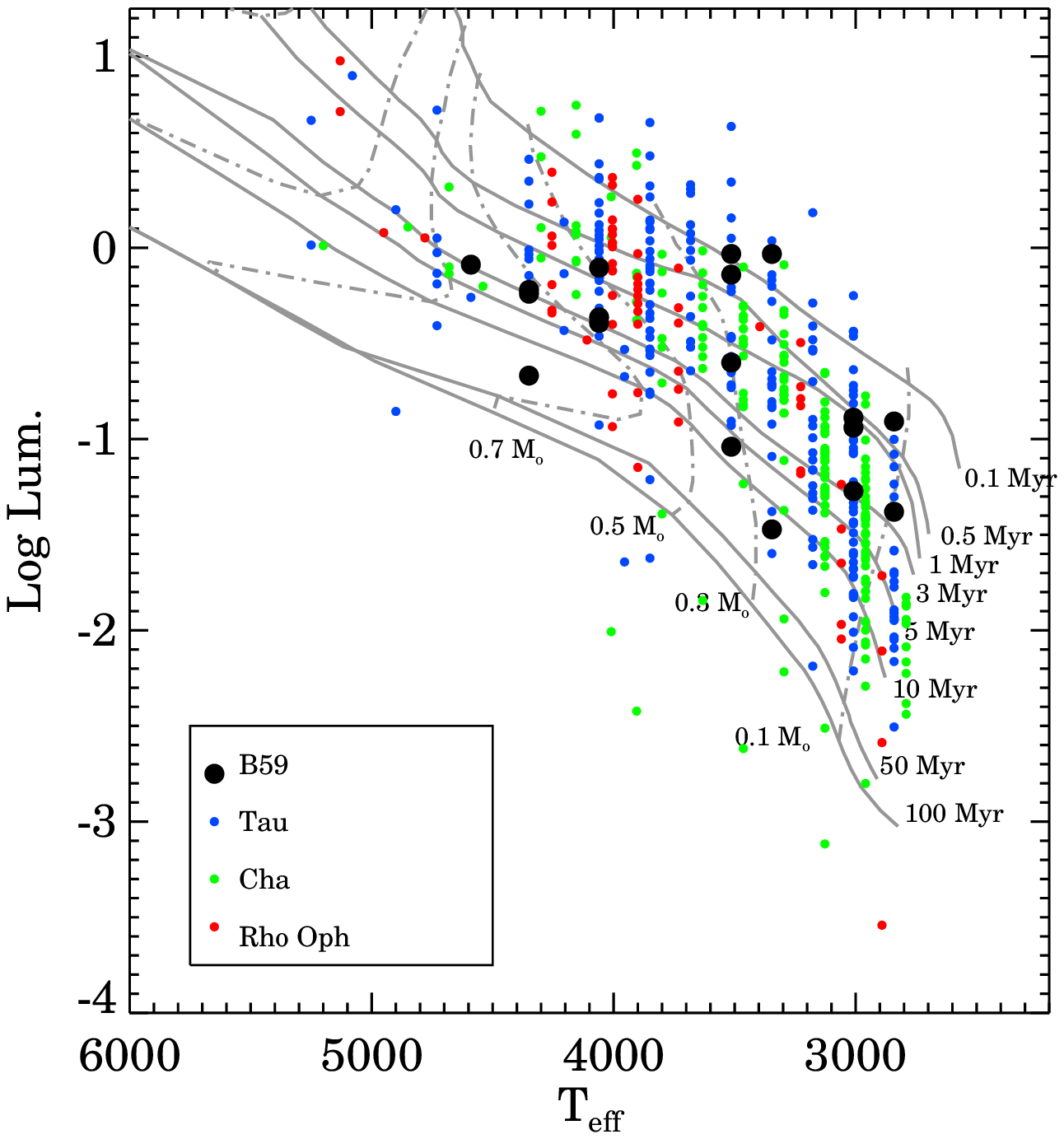}
\caption{ \normalsize{HR diagram locations for YSOs in nearby young clusters, calculated using JHK photometry and assuming stellar luminosities (subtracting off H band veiling).  B59 sources are shown as black dots, Rho Ophiuchus sources are shown in red, Taurus sources are shown in blue, and Chameleon sources are shown in green.  The DM98 model grid is shown for comparison with solid and dotted lines corresponding to isochrones and fixed mass evolutionary tracks, respectively.}}\label{fig:compareHRs}
\end{figure*}

We also measured median stellar ages for other, 
well studied young clusters to provide an indication of B59's
relative, rather than absolute, age.  These cluster age measurements 
require YSOs catalogs with measured spectral types and JHK
photometry to estimate extinction and veiling corrected
stellar luminosities using the same algorithms we applied to B59's YSOs.
Studies by Luhman and collaborators provide such catalogs for YSOs in 
Taurus, Rho Ophiuchus, and Chameleon \citep{Luhman1999,Luhman2006b,Luhman2007}.
We have placed these objects on the HR diagram using the same routines
applied to our B59 data, adopting the following distances to each region: Taurus -- 140 pc., Oph -- 126 pc., Cha -- 170 pc \citep{Kenyon2008,Wilking2008,Luhman2008}.  Table \ref{tab:cluster_ages} 
presents the median stellar age implied for each cluster by 
the DM98 and \citet{Baraffe1998} PMS tracks, and
Figure \ref{fig:compareHRs} shows the locations of each region's YSOs
in the HR diagram.  

\begin{table*}
\begin{center}
\caption{Median Cluster Ages}
\begin{tabular}{l | c | c c | c c}
\hline \hline
Cluster & N$_{stars}$ & DM98 Median &  B98 Median & Statistical & Systematic \\
        &             & Age (Myrs) & Age (Myrs) & Uncertainty (1$\sigma$) & Uncertainty \\
\hline
B59 & 18 & 2.59 & 4.37 & 29\% & 100--175\% \\
Taurus & 224 & 1.53 & 2.63 & 9\% & 100--175\% \\
$\rho$ Oph & 61 & 1.04 & 3.63 & 16\% & 100--175\% \\
Chameleon & 185 & 2.59 & 3.02 & 9\% & 100--175\% \\
\hline
\end{tabular}
\label{tab:cluster_ages}
\end{center}
\end{table*}

Our analysis of the ages of YSOs in other regions, however, lacks the iterative refinement of each YSO's veiling estimate according to the agreement between the strengths of spectral features in the target and artificially veiled standard spectrum.  Instead, for YSOs in other regions, we simply adopt the mean CTTS J band veiling measured by Cieza et al. (2005), which implicitly assumes that CTTSs dominate the population of each star forming region.  This J band excess is an overestimate of the minimal veiling flux detected from WTTSs, however, such that we will under-estimate WTTSs intrinsic stellar luminosity, and thus overestimate their age.  Our age estimates for Taurus, $\rho$ Oph and Chameleon will 
therefore be slightly biased towards older ages in proportion to the number of 
WTTSs in the sample: we expect this to be a minimal effect for Taurus and $\rho$ Oph, but
potentially of some importance for the age we measure for the Chameleon star forming region.

\subsection{Uncertainties in Derived Ages}
\label{ageerrors}

Our measurement of B59's median stellar age is subject to 
a few distinct statistical and systematic uncertainties. 
These affect the relative and absolute
precision of our measurement in different ways, so we discuss
them each below.

\subsubsection{Random/Statistical Age Uncertainties}

Our measurement of cluster median stellar ages are 
susceptible to (presumably) random uncertainties due
to observational error and sampling effects.  
Age uncertainties introduced
by observational errors are straightforward: errors in a YSO's temperature and luminosity
will shift its location in the HR diagram, corresponding to an error
in the inferred mass and/or age.  In the temperature/luminosity regime occupied by B59's YSOs, luminosity is primarily sensitive to a star's age, while its temperature is a better indicator of its mass.  We therefore estimate the age errors 
that could be introduced by uncertainties in our luminosity estimates: we find that
a 0.3 dex luminosity error corresponds to an age error of $\sim$125\%.  This is, however, the 
error in any individual age estimate: as we are determining the median stellar age of all the YSOs in 
each cluster, we can expect that this error will scale like $\sqrt{N_{cluster}}$.  We calculate
an estimate of the statistical uncertainty in each cluster's median stellar age as
$\frac{125\%}{\sqrt{N_{cluster}}}$, and present the resulting uncertainty in  Table \ref{tab:cluster_ages}.

Sampling errors are perhaps the more subtle effect, but could be important for a sparse 
cluster like B59.  A simple demonstration
of this effect can be seen by examining the ages inferred for the 18 B59 
YSOs listed in Table \ref{tab:inferred_params}.  The median age of these 
B59 YSOs, as inferred from the DM98 models, is bounded by the 2.42 Myr and 
2.59 Myr ages estimated for V359 Oph and [BHB2007] 3, respectively the 
9$^{th}$ and 10$^{th}$ oldest YSOs in the list.  The relatively narrow 
age range bounding the median, however, is somewhat misleading.  There are at least
three B59 YSOs that were too heavily embedded to be observationally accessible
to SpeX: [BHB2007] 10 and 11, classified as Class 0/I and Class I sources by \citet{Brooke2007}, 
and 2M17112255, discovered and classified as a Class I system by \citet{Riaz2009}.
These YSOs are presumably quite young: arbitrarily assigning them ages of 0.5 Myrs would
shift the age ordering such that the YSO defining B59's median stellar age would be 
[BHB2007] 20, with an age of 1.37 Myrs according to the DM98 models.  This would 
represent a $\sim$50\% age error, somewhat larger than the 30\% age estimate 
predicted by the simple $\sqrt{N_{cluster}}$ calculation 
described above.  We lack, however, a robust statistical description of the effect, so merely
caution the reader that the formal statistical uncertainties reported in Table \ref{tab:cluster_ages} 
may be underestimates for a sparse cluster like B59.

\subsubsection{Systematic Age Uncertainties: Veiling Corrected Luminosities} 

One systematic effect which could affect our age estimates concerns the treatment of veiling 
flux in calculating each YSO's luminosity.  Deriving ages from each YSO's 
total luminosity, rather than the veiling corrected stellar luminosity, 
typically reduces the YSO's implied age by $\sim$1.5 Myrs over the 
range of ages considered here.  The median stellar age 
implied for B59 by comparing each YSO's total, non-veiling corrected 
luminosity estimate to the DM98 tracks is 0.87 Myrs, as opposed to the 
veiling corrected age estimate of 2.6 Myrs.  This represents a nearly 
75\% decrease in the cluster's median stellar age, demonstrating
that correctly accounting for veiling luminosity is necessary
to accurately infer a YSO's age.  As the PMS tracks considered here
do not account for emission from accretion or a circumstellar disks,
we believe the correct approach is to infer ages from veiling-corrected 
luminosity estimates.  We caution the reader, however, that many other 
investigators adopt non-veiling corrected bolometric luminosity estimates 
to derive YSO ages: as demonstrated above, this difference in approach 
results in age differences of 50-100\%, and should be carefully 
considered in reconciling our results with other studies in the literature.

\subsubsection{Systematic Age Uncertainties: Adopted PMS Tracks} 

As shown by \citet{Hillenbrand2008}, the choice of PMS models from which
to infer YSO ages has a significant affect on the absolute age derived
for a given cluster.  \citet{Hillenbrand2008} found that the 
PMS models computed by DM98 and \citet{Baraffe1998} predict the 
youngest and oldest ages, respectively, for a YSO of a given 
temperature and luminosity.  We therefore use these two sets 
of model tracks to span the full range of B59's potential 
absolute age.  Using the same veiling corrected luminosities
derived above for B59's YSOs, but inferring ages from
the $\alpha=1$ PMS tracks calculated by \citet{Baraffe1998}, results 
in a median stellar age for B59 of 4.4 Myrs, or roughly 70\% 
older than the median stellar age inferred from the DM98 models. 

Model dependent age differences are often attributed to luminosity 
offsets between equal age isochrones in two sets of PMS tracks.  
This suggests that, while the absolute ages predicted
by two sets of PMS tracks may differ, they should produce similar \textit{relative}
age orderings for a given sample of clusters.  Table \ref{tab:cluster_ages} 
reveals, however, that this simple expectation is not necessarily 
borne out in practice: both sets of models indicate that B59 has the 
oldest median stellar age, but the DM98 models imply that $\rho$ Oph is 
\textit{younger} than Taurus and Chameleon, while the 
\citet{Baraffe1998} models identify $\rho$ Oph as 
\textit{older} than Chameleon and Taurus.  

\begin{figure*}
\epsscale{1.0}
\plotone{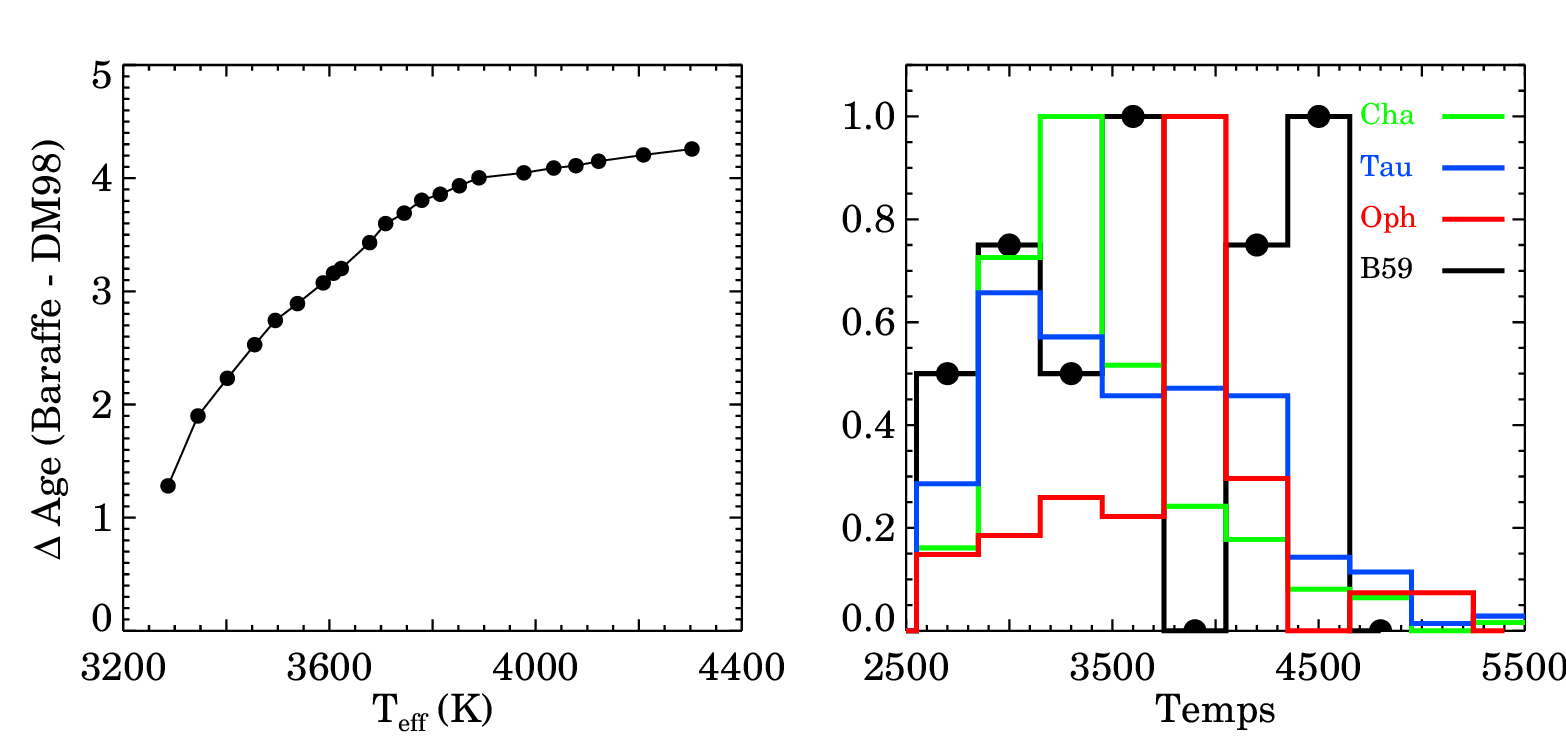}
\caption{ \normalsize{(Left panel) The age difference between a 5 Myr Baraffe isochrone and the age indicated by the DM98 models for the same location in the HR diagram, presented as a function of T$_{eff}$.  (right panel) T$_{eff}$s of the stars sampled from each cluster analyzed here.  The Cha \& Taurus samples are dominated by cooler stars, where the disparity between the B98 and DM98 models is $\sim$1 Myr.   B59 and $\rho$ Oph are successively warmer: at these temperatures, the DM98 and B98 models predict ages that can differ by 4 Myrs. }}\label{age_order_fig}
\end{figure*}

These different relative age orderings arise from differences in the
stellar population sampled in each cluster, combined with distinct isochrone slopes
predicted by the two sets of PMS tracks.  The temperature dependent offset between
the 5 Myr DM98 and Baraffe isochrones is shown in the left panel of Figure \ref{age_order_fig}.
The Baraffe models imply consistently older ages for a given HR diagram location, but the age 
difference increases rapidly towards higher T$_{eff}$.  This means that the relative age orderings 
of single stellar populations may not be preserved across models if the YSOs observed 
sample different temperature ranges.  This is precisely the cause of the different relative ages inferred for
$\rho$ Oph in the DM98 and Baraffe models:as demonstrated in the right panel of Figure \ref{age_order_fig}, the median temperature of the $\rho$ Oph cluster stars is significantly larger than that of the other clusters analyzed here.

\subsubsection{Systematic Age Uncertainties: Cluster Distances} 

Uncertainties in a cluster's distance can also introduce errors in 
the inferred age: a closer distance implies a lower luminosity, and 
thus an older age, for a YSO of a given apparent magnitude. 
A concrete example of of this effect is provided by 
$\rho$ Oph, whose distance has recently been revised to $\sim$125 
parsecs from a previous estimate of $\sim$160 parsecs.  The 0.3 Myr 
estimate of $\rho$ Oph's age derived by \citet{Luhman1999} was based
on the 160 parsec distance estimate.  Replicating their analysis with a revised distance
 of 125 parsecs increases the derived median stellar age to 0.63 Myrs, twice
as old as the previous estimate. Estimates of the distance to B59 and/or the Pipe 
Nebula are scarce: the two most recent distances measured for the Pipe 
are 130$^{+13}_{-20}$ parsecs and 145$\pm$16 parsecs, obtained by \citet{Lombardi2006} 
and \citet{Alves2007a}, respectively.  We have adopted the
\citet{Lombardi2006} distance for the luminosity and age measurements
presented here, but we note that the \citet{Alves2007a} 
distance implies in a median stellar age (inferred from veiling corrected luminosities and
DM98 tracks) for B59 of 1.7 Myrs, or decreased by roughly 35\% from our primary
age estimate.

\subsubsection{Total Error Budget for B59's Age Estimate} 
In summary, our measurement of B59's absolute age is subject to statistical errors at the 
$\sim$30\% level, such that B59's median stellar age (adopting veiling corrected luminosities 
and inferred from the DM98 models) is 2.6$\pm0.8$ 
Myrs.  Several systematic uncertainties could shift the median stellar 
age derived for B59, however, by as much as 100\%. Adding these effects in quadrature, we find 
that systematic uncertainties could introduce errors as large as 150\% into our 
measurement of B59's absolute age.  Including potential systematic effects therefore revises the 
error budget for B59's median (DM98) stellar age to 2.6$^{+4.1}_{-2.6}$ Myrs.  

\section{Discussion}

\subsection{The Star Formation Timescale in B59}

The prompt star formation paradigm agrees well with the 
evolutionary timescales expected for dense cores that lack significant internal or external support.  Acted on by gravity alone, a core should collapse 
in roughly one free-fall time, $\tau_{ff} = (3\pi/ 32G\rho)^{1/2}$. 
Hydrodynamic models of undriven turbulent molecular clouds predict only 
slightly longer evolutionary timescales, as the turbulent support 
dissipates on $\sim$1-2 free-fall timescales \citep{Klessen2000}. \citet{Offner2008}, for example, 
find that star formation activity ceases after 0.75 $\tau_{ff}$ in their 
simulations of star formation within undriven turbulent molecular cores.

The uncertainties associated with our measurement of B59's age 
are sizable, but the most straightforward
interpretation of this result is that star formation activity began 
in B59 nearly 3 Myrs ago.  The presence of heavily embedded (and presumably
young) Class 0/I objects within B59 \citep{Brooke2007,Riaz2009} also
implies that this star formation activity has persisted to the 
present day.  If the oldest stars in B59 formed from the dense core 
currently associated with the cluster, these observations would indicate 
a lower limit of 3 Myrs for the core's lifetime.  A star formation timescale 
for B59 comparable to 3 Myrs is by no means startling in absolute terms.  
Indeed, while the stellar content of B59 appears somewhat older than 
that of the $\rho$ Oph, Taurus, and Chameleon star forming regions, 
the inferred star formation timescales agree to within a factor of two.  

Expressing B59's age in units of the core free-fall time and the 
core's crossing time is also useful for interpreting the core's physical evolution.  Adopting the 
density (n(H$_2$) = 10$^4$ cm$^{-3}$; $\rho = 3.5\times10^{-20}$ g cm$^{-3}$), 
radius (R = 0.19 parsecs = 5.9$\times10^{12}$ km), and velocity dispersion 
($\sigma_v$=0.37 km/sec, calculated from 0.86 km/sec FWHM) 
tabulated for the B59 dense core 
by \citet{Rathborne2009}, we calculate $\tau_{ff} \sim$0.35 Myrs and 
$\tau_{cross} \sim$0.5 Myrs.  These calculations describe B59's current 
state; if B59 is evolving from a lower density to higher density 
configuration, these timescales would have been somewhat longer in 
the past than they are today.  Intensive observations of the dense gas 
within B59 have not revealed any kinematic 
signatures of rapid collapse \citep{Rathborne2009}.  If B59 is 
evolving relatively quiescently, the timescales we derive
today are likely comparable to those of B59's recent past.

The uncertainties on B59's age permit that star formation
may have begun as recently as the core's last dynamical time, or as long
as $\sim$13 dynamical timescales ago. The maximum likelihood
value, however, suggests that B59 has been forming stars for 
$\approx$6 crossing or free fall times.  Extinction maps of B59 
also indicate that the core is spatially coherent, having not
undergone significant fragmentation as it formed stars \citep{RomanZuniga2009}.  
Taken at face value, the maximum likelihood value of B59's star formation timescale
suggests that B59 could join a short, but growing, list of dense cores that appear to 
have resisted global collapse and sustained ongoing 
star formation for several dynamical times. \citet{Swift2008a} examined the dense gas and stellar 
population of the L1551 star forming region, 
demonstrating that L1551 has been forming stars for $\sim$5 Myrs,
or $\geq$5 $\tau_{ff}$; they identify stellar feedback as a likely 
source of support against collapse within this region. As we find 
for B59, \citet{Swift2008a} also find that L1551 contains stars 
over a broad range of evolutionary stages (Class 0 through III). 
Similar results have been reported for the young cluster AGFL961; 
\citet{Williams2009} conclude that either disk evolution proceeds 
more rapidly in clustered environments, or that star formation
persists in clusters over several dynamical timescales. 
Finally, \citet{Tan2006} present observationally-based calculations that 
suggest ages of $\sim$3-4 $\tau_{ff}$ for several rich young 
clusters that are still forming stars (ie, $\rho$ Oph, IC 348, ONC, etc.). 
These measurements suggest that some dense cores can sustain star
formation over a few dynamical timescales, including in relatively low-mass
cores like B59, which does not appear to host massive stars.

What physical mechanism could be supporting these cores against
collapse, extending their lifetimes beyond that expected for cores in
free fall collapse?  As noted above, one possibility is stellar
outflows: \citet{Matzner2007} calculate that outflows may support 
clusters against collapse for several crossing times.  This mechanism could explain B59's relative stability, particularly given that 
\citet{Onishi1999} and \citet{Riaz2009} find evidence for an active outflow within the B59 core.
A recent ammonia survey of the Pipe, however, finds little evidence for significant non-thermal motions within B59's dense gas, as would be expected if bulk outflow motions
are the core's main support against gravitational collapse \citep{RomanZuniga2009,Rathborne2008}. 
High-resolution observations of the kinematics of
B59's dense gas are required to clarify the extent to which it (and, by 
extension, outflows from previous generations of protostars) could
support B59 against collapse. 

The collapse of a molecular core may also be strongly effected by 
radiative feedback from the young protostars it contains.  Krumholz, Klein \& McKee (2007) used simulations of massive molecular cores to demonstrate that protostars can contribute significant heating to a collapsing core, thereby suppressing fragmentation.  This heating does not, however, appear to slow mass accretion and collapse within the core: much of the radiation `leaks' out of optically thin holes in the core (Krumholz et al. 2009).  Offner et al. (2009) extended this investigation into the regime of low-mass star formation, finding a similar reduction in stellar multiplicity due to reduced fragmentation, as well as a significant reduction in the mass accretion rate of the earliest formed protostar.  These simulations suggest that radiative feedback could play some role in extending B59's star formation timescale.

Magnetic fields are a final physical mechanism that is often invoked to support cloud cores
against collapse.  Magnetic fields have been a 
fundamental component of star formation models for decades 
\citep{Nakano1984,Shu1987,Mouschovias1999}, but observational
constraints on the strengths of magnetic fields within molecular clouds 
are difficult to obtain \citep[e.g.,][]{Crutcher2009}.
Past measurements suggest magnetic fields
may contribute appreciable amounts of support to molecular
cores \citep{Troland2008}, and recent polarization measurements
in the Pipe Nebula indicate the presence of a $\sim$17 $\mu$G magnetic
field within B59 itself \citep{Alves2008}.  This suggests that magnetic fields may be providing B59 with 
some support against collapse, but this is a relatively weak statement: magnetic fields 
can extend a dense core's lifetime arbitrarily, depending on the field strength adopted.  

\subsection{The Star Formation Efficiency of B59}

As the discussion above demonstrates, there are several mechanisms which 
could plausibly explain B59's maximum likelihood star formation timescale.  To provide additional
leverage for comparing observations of B59 with predictions of theoretical
models of star forming cores, we now consider B59's star formation efficiency 
in addition to its characteristic timescale.
 
B59's present-day star formation efficiency (SFE) can be
characterized as the mass of the cluster's stellar population divided 
by the core's initial gas mass. 
The total mass of all B59 YSOs analyzed here is $\sim$8 M$_{\odot}$ according to the DM98 models, and 12 M$_{\odot}$ according to the \citet{Baraffe1998} models. \citet{Brooke2007} and \citet{Riaz2009} 
also provide preliminary mass estimates for [BHB2007] 10, 11, and 2M12112255, 
the B59 YSOs that were too heavily embedded to be observed with SpeX.  These
embedded YSOs contribute an additional 0.7 M$_{\odot}$ to B59's total stellar mass, 
bringing B59's total stellar mass to $\sim$9-13 M$_{\odot}$.

B59's initial gas mass can be approximated as the sum of its stellar and gas
masses.  Extinction maps indicate that B59 contains $\sim$20 M$_{\odot}$ of dense gas,
implying an initial core mass of $\sim$29-33 M$_{\odot}$.  This suggests
B59's present-day SFE is $\sim$30-40\%, depending on
the choice of models used to infer stellar masses.  Combining this SFE
with B59's median stellar age produces an estimate of the SFE/$\tau_{ff}$ of the B59 dense core.
 B59's sizable age uncertainty translates into a large range of allowed SFE/$\tau_{ff}$ values: 3-40\%, though 
the probability distribution is strongly asymmetric, peaking at a maximum likelihood 
value of $\sim$6\%.

The SFE/$\tau_{ff}$ values predicted by theoretical models for star forming cores cover
a large range, with slower star formation rates corresponding
to models incorporating some additional physical mechanism (driven
turbulence, magnetic fields, outflows, etc.) to support cores 
against collapse.  Unsupported models typically produce SFE/$\tau_{ff}$ values
$\sim$ 10-20\%, while models that incorporate additional support often 
predict proportionally lower SFE/$\tau_{ff}$ values.  
As one example, the moderately supercritical (ie, prone to 
gravitational collapse) magnetic models 
of \citet{Vazquez-Semadeni2005a} produced SFE/$\tau_{ff}$ values
$\sim$ 5\%, whereas earlier non-magnetic models by the same
authors produced SFE/$\tau_{ff}$ values $\sim$15\% \citep{Vazquez-Semadeni2003}.  The 
magnetically subcritical models of \citet{Nakamura2008}, which
also incorporate feedback from stellar outflows, produce
SFE/$\tau_{ff}$ values of 1\% or less.

The theoretical models calculated by \citet{Price2009} are 
arguably the best match to B59's actual physical structure.
These models (indicated hereafter as `the PB09 model grid') 
incorporate both magnetic fields and radiative feedback, and follow the evolution of a cloud 
core whose mass (50 M$_{\odot}$) and radius ($\sim$0.2 pc) are reasonably 
close to the values which likely characterized B59 prior to the onset 
of active star formation.  The PB09 model grid spans magnetic fields
of various strengths, but all models were supercritical, with 
mass-to-flux ratios ranging from 3 to $\infty$.  Though the models
were only allowed to evolve over a timespan corresponding to 1.5 $\tau_{ff}$,
or 0.35--0.5 $\tau_{ff}$ following the formation of the first star in the core,
the implied SFE/$\tau_{ff}$ values range from $\sim$30\% for the model
lacking both magnetic fields and radiative feedback, to 
$\sim$ 6\% for the model incorporating radiative feedback and the 
strongest magnetic field (B$_0$=65$\mu$G).  In analyzing the range
of star formation efficiencies produced by their model grid, PB09 note that
magnetic fields appear to play a more important role than radiative feedback 
in quenching star formation.

B59's maximum likelihood SFE/$\tau_{ff}$ value of $\sim$6\% is contained within the range of 
values predicted by the PB09 models, but B59's observed magnetic field
strength (B$\sim$17 $\mu$G) is only $\sim$1/4 that adopted in the least efficient PB09
model (B$\sim$65 $\mu$G).  The PB09 models and
the present-day B59, however, represent significantly different time periods in the 
evolution of a star forming core. It would be interesting to see what properties the 
PB09 models might have if they were allowed to evolve to $\sim$5 $\tau_{ff}$: would the 
magnetic field diffuse out of the core, such that the measured magnetic field strength would 
better agree with what we observe in B59 today?  

It is too early to draw detailed conclusions, but the agreement between B59's observed
properties and the predictions of the PB09 model grid are intriguing, and suggest
that magnetic support and radiative feedback help support B59 against collapse.  

\subsection{The Spatial Extent of Star Formation in Pipe}

The suggestion that magnetic fields may be important for understanding
star formation within the Pipe Nebula is also consistent with the results of 
the polarization survey conducted by \citet{Alves2008}.  \citet{Alves2008} measure a significant 
spatial gradient in the component of the Pipe's magnetic field aligned
with the plane of the sky, ranging from $\sim$17 $\mu$G in B59 to 
$\sim$65$\mu$G in the Pipe's `Bowl'.  As \citet{Alves2008} indicate, this gradient
correlates with star formation activity: the weakest, most poorly 
aligned field vectors lie in B59, where the bulk of the Pipe's star formation 
activity is taking place, while areas with stronger, better aligned magnetic 
field vectors appear devoid of star formation activity.  

In this context, it is interesting to note the location of [CLR2010] 2, the 
candidate YSO identified by \citet{Forbrich2009} in their
MIPS survey of the Pipe, and subsequently confirmed as a likely YSO
by our spectroscopic analysis.  This candidate YSO lies in the Pipe's 
`Stem', where the magnetic field appears to be stronger than in 
B59, but not as strong as in the Bowl.  This YSO also displays a 
rich emission line spectrum, suggesting it is a young, heavily 
accreting YSO, which would be consistent with a picture in which
star formation activity was just beginning in the Neck of the Pipe.
We note, however, that [CLR2010] 2 does not appear to be 
heavily embedded; indeed, [BHB2007] 10 and 11 appear to be in
significantly earlier evolutionary stages than [CLR2010] 2, making
it difficult to conclude that the onset of star formation activity 
is simply progressing linearly from B59 towards the Bowl of the Pipe.

\section{Summary}

\begin{enumerate}
\item{We obtained and analyzed JHK SpeX spectra for 20 candidate B59 YSOs, and 10 additional mid-IR bright sources throughout the larger Pipe Nebula.  We measured temperature and luminosity sensitive spectral indices from these spectra, and determined the extinction, NIR color excesses, and spectral type required to self-consistently reproduce each target's spectral energy distribution.  This analysis leads us to re-classify [BHB2007] 5 and 17, previously identified as candidate B59 members, as background giants, and to identify [CLR2010] 2 SW as a promising candidate YSO in the stem of the Pipe. None of the mid-IR bright candidates outside of the Stem of the Pipe were found to be YSOs}
\item{For each candidate B59 YSO with reliable spectra and photometry, we derived effective temperatures and luminosities, with and without correction for non-photospheric flux.  Using these measurements to place each YSO in the HR diagram, we inferred ages and masses for each YSO from theoretical pre-main sequence evolutionary tracks. B59's median 
stellar age, as inferred from the DM98 models and accounting
only for statistical uncertainties, is 2.6$\pm$0.8 
Myrs.  Including potential systematic effects revises the 
error budget for B59's median (DM98) stellar age to 
2.6$^{+4.1}_{-2.6}$ Myrs.}
\item{Comparing the median stellar ages implied for several star forming regions by different pre-main sequence evolutionary tracks demonstrates that the construction of even a relative age ordering requires careful attention to sampling of each region's stellar population.  Despite this sensitivity, both sets of tracks studied here imply that B59 is at least as old as, if not slightly older than, the more well-studied $\rho$ Oph, Taurus, and Chameleon star forming regions. The dominant source of error in this conclusion is the uncertainty in the distance adopted to each cluster: we minimize other potential systematic errors by applying a consistent set of algorithms to a homogeneous set of catalogs of young stars in each cluster.}
\item{The maximum likelihood median stellar age we measure for B59, and the observed properties of the dense gas in the region, suggests that the B59 dense core may have been stable against collapse and actively forming stars for roughly 6 dynamical timescales, with a star formation efficiency per dynamical time of $\sim$6$\%$, though the uncertainty in our age measurement permits a significant range of possible SFE/$\tau_{ff}$ values (3--40\%).  These properties agree well with recent star formation simulations that incorporate sub-critical magnetic fields and radiative feedback from protostellar heating. }
\end{enumerate}

\acknowledgements The authors owe a debt of gratitude to Bill Golisch, Paul Sears, and Bobby Bus for their generous assistance in obtaining efficient and accurate observations.  KRC gratefully acknowledges Kevin Luhman's assistance in providing
homogeneous catalogs of spectral types and photometry for young stars
in well known star forming regions, and helpful discussions with Lynne Hillenbrand and Michael Meyer concerning the practice and reliability of pre-main sequence age measurements, with Eric Mamajek on the history of star formation in the Sco-Cen region, and Stella Offner on the physics of collapsing cores.  The authors also wish to recognize and acknowledge the very significant cultural role and reverence that the summit of Mauna Kea has always had within the indigenous Hawaiian community.  We are most fortunate to have the opportunity to conduct observations from this mountain.  NASA support was
provided to K.\ Covey for this work through the Spitzer Space Telescope Fellowship
Program, through a contract issued by the Jet Propulsion Laboratory,
California Institute of Technology under a contract with NASA. 

This research has made use of NASA's Astrophysics Data System
Bibliographic Services, the SIMBAD database, operated at CDS,
Strasbourg, France, the NASA/IPAC Extragalactic Database, operated by
the Jet Propulsion Laboratory, California Institute of Technology,
under contract with the National Aeronautics and Space Administration,
and the VizieR database of astronomical catalogs
\citep{Ochsenbein2000}. IRAF (Image Reduction and Analysis Facility) is
distributed by the National Optical Astronomy Observatories, which are
operated by the Association of Universities for Research in Astronomy,
Inc., under cooperative agreement with the National Science Foundation.

The Two Micron All Sky Survey was a joint project of the University of
Massachusetts and the Infrared Processing and Analysis Center
(California Institute of Technology). The University of Massachusetts
was responsible for the overall management of the project, the
observing facilities and the data acquisition. The Infrared Processing
and Analysis Center was responsible for data processing, data
distribution and data archiving.

\appendix

\section{TWA Spectral Standards}\label{app:TWHyaatlas}

\begin{figure}
\epsscale{0.75}
\plotone{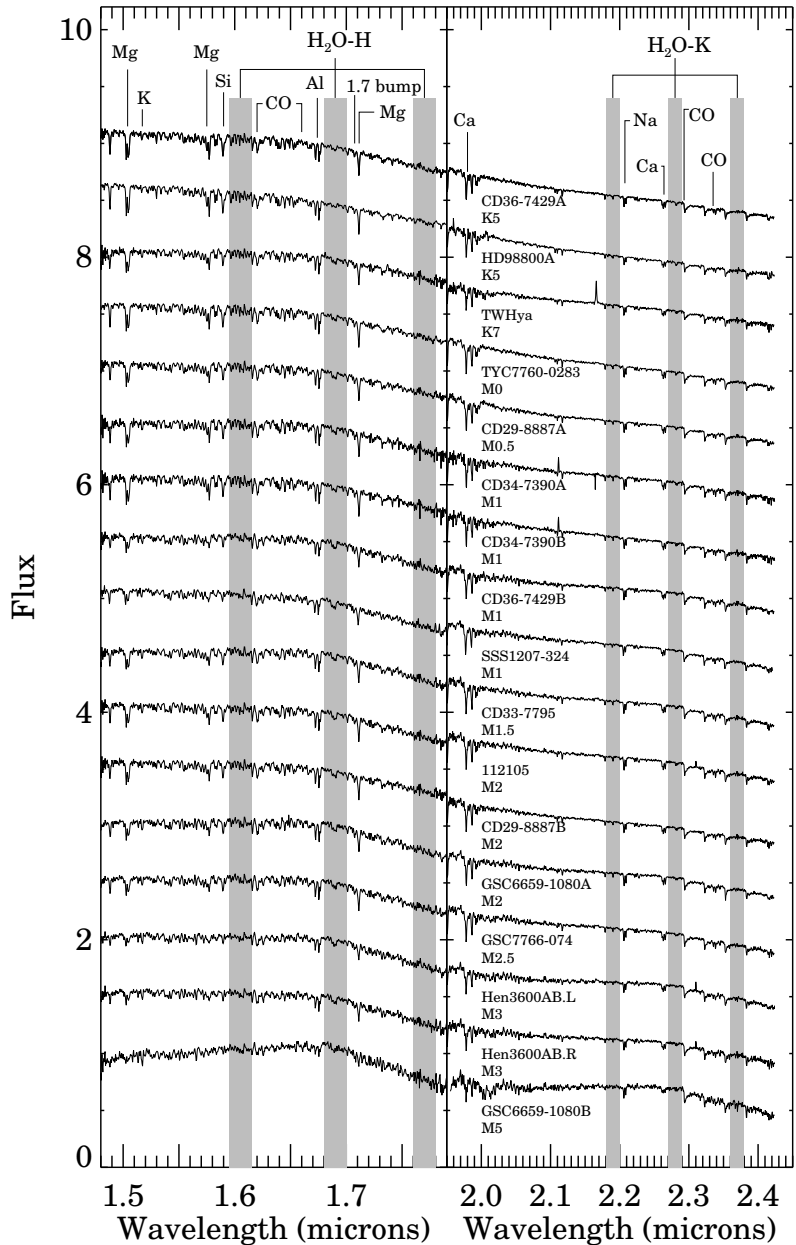}
\caption{ \normalsize{HK bands of TWHya standard spectra.  }}\label{fig:TWHyaHK}
\end{figure}

To supplement the spectral atlas presented by \citet{Cushing2005} with
additional pre-main sequence spectral standards, we obtained
SpeX SXD spectra of several members of the TW Hya moving group.  Observed 
during the same nights as our B59 candidates, these spectra share
the same instrumental set-up and reduction process as our target
spectra. The H and K band portion of these spectra are shown in Figure
\ref{fig:TWHyaHK}, ordered such that earlier type objects are found higher 
in the figure.  Spectral features sampled by the indices presented in \S 
\ref{measurespt} are also labeled.  

\begin{figure}
\epsscale{0.95}
\plotone{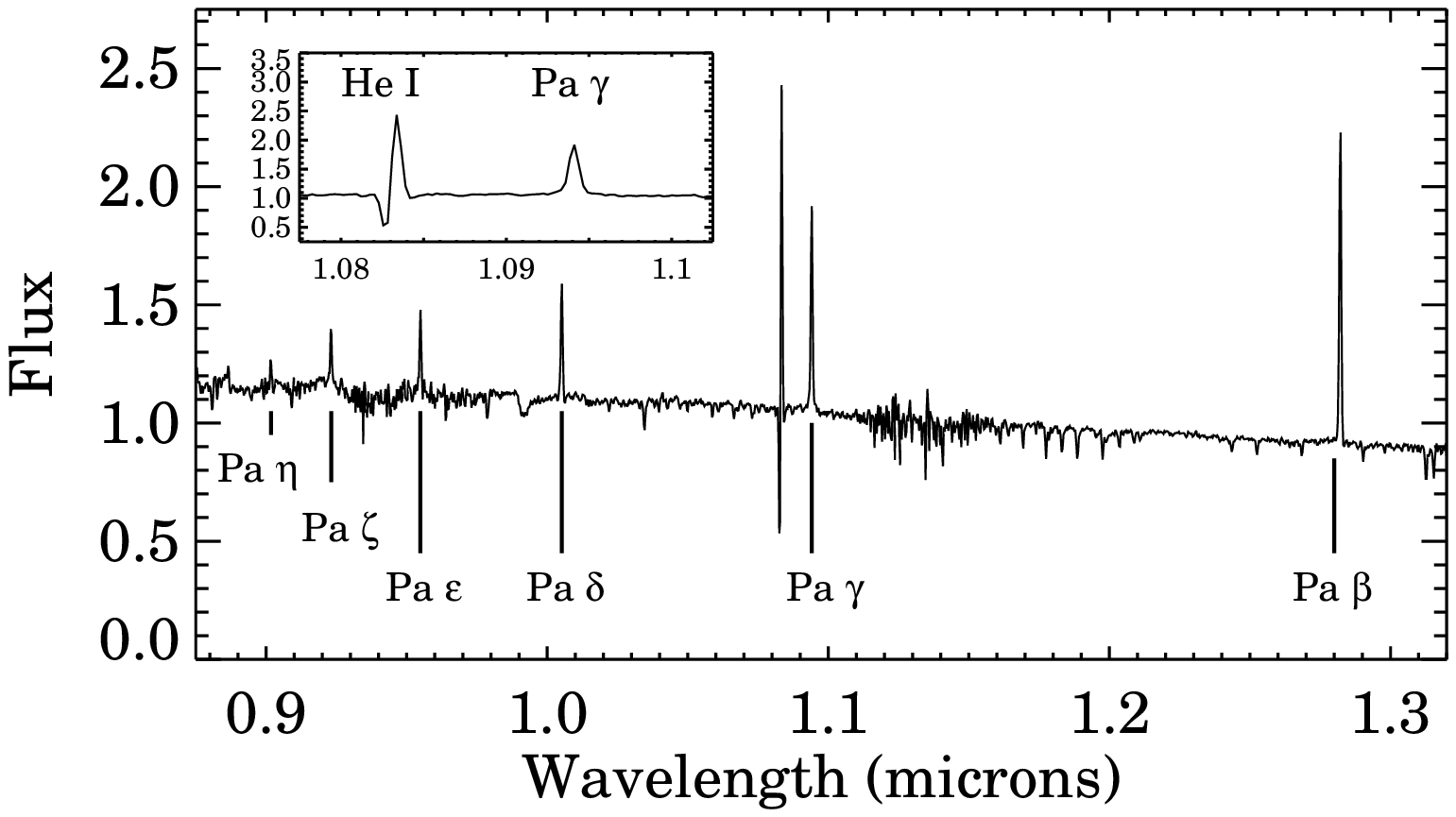}
\caption{ \normalsize{J band spectrum of TW Hydra to show rich emission line spectrum.}}\label{fig:TWHyaJ}
\end{figure}

As most B59 members are heavily reddened, the
z and J band portion of their observed spectrum was typically 
quite noisy, and thus not useful for measuring spectral types.  
This region contains most of the HI Paschen series, the 
Ca II triplet and the He I 10830~\AA~line, however, which are
often seen in emission in young, actively accreting stars. 
The 0.8-1.3 $\mu$m SpeX spectrum of TW Hya, an actively
accreting CTTS, is shown in Figure \ref{fig:TWHyaJ} for comparison
with Figure \ref{fig:tarspecs-J}, which shows 0.8-1.3 $\mu$m spectra
of candidate or confirmed B59 members with J band emission features.

To provide the community with moderate resolution NIR pre-main sequence spectral standards, we have made these spectra publicly available at \url{http://www.astro.cornell.edu/~kcovey/CoveyB59SpeX.tar.gz}.

%--------------------------BIBLIOGRAPHY---------------------------
%\clearpage
\setlength{\baselineskip}{0.6\baselineskip}

\setlength{\baselineskip}{1.667\baselineskip}

\end{document}

%% file: t1.tex
\begin{deluxetable*}{lcccccccccccc}
\tablewidth{0pt}
\tabletypesize{\tiny}
\tablecaption{IRTF/SpeX SXD Observing Log for Candidate B59 Members \label{tab:obslog_B59}}
\tablehead{
  \colhead{Source} &
  \colhead{} &
  \colhead{} &
  \colhead{2MASS} &
  \colhead{2MASS} &
  \colhead{2MASS} &
  \colhead{SXD} &
  \colhead{SXD} &
  \colhead{SXD} &
  \colhead{SXD} &
  \colhead{SXD} &
  \colhead{NTT} &
  \colhead{NTT}\\
  \colhead{ID} &
  \colhead{RA} &
  \colhead{Dec} &
  \colhead{$J$} &
  \colhead{$H$} &
  \colhead{$K_{s}$} &
  \colhead{itime} &
  \colhead{slit} &
  \colhead{J S/N} &
  \colhead{H S/N} &
  \colhead{K S/N} &
  \colhead{$H$} &
  \colhead{$K$} }\startdata
\textrm{[BHB2007]} 1 &  17:11:03.93 & -27:22:55.08 &  10.46 &   8.99 &   7.76 &   240 &     0.3 &  152/52 &   243/49 &  293/114 & 9.29 & 8.29\\
\textrm{[BHB2007]} 2 &  17:11:04.12 & -27:22:59.44 &   9.75 &   8.79 &   8.05 &   240 &     0.3 &  136/48 &   155/35 &   151/73 & 8.91 & 8.56\\
\textrm{[BHB2007]} 3 &  17:11:11.83 & -27:26:54.98 &  14.09 &  12.62 &  11.76 &  6240 & 0.5/0.3 &   58/13 &   108/26 &   186/40 & 12.69& 11.77\\
\textrm{[BHB2007]} 4 &  17:11:14.46 & -27:26:54.45 &  11.62 &  10.42 &   9.63 &  1920 &     0.3 &  277/41 &   344/48 &   347/57 & 10.62& 9.77 \\
\textrm{[BHB2007]} 5 &  17:11:16.26 & -27:20:28.82 &  11.23 &   9.70 &   9.07 &  1440 &     0.3 &  132/57 &   168/23 &   158/56 & 9.70& 9.08 \\
\textrm{[BHB2007]} 6 &  17:11:16.31 & -27:25:14.50 &  10.58 &   9.37 &   8.75 &   960 &     0.3 &  203/40 &   234/41 &   227/63 & 9.37& 8.75 \\
\textrm{[BHB2007]} 7 &  17:11:17.28 & -27:25:08.16 &  13.61 &  10.82 &   8.77 &  1920 &     0.3 &  58/12  &   261/34 &   325/92 & 10.58& 8.77 \\
\textrm{[BHB2007]} 8 NW & 17:11:18.24&-27:25:48.94 &\nodata & \nodata & \nodata &  1920 &     0.3 & \nodata &      2/0 &    59/17 & 16.01& 12.87 \\
\textrm{[BHB2007]} 8 SE & 17:11:18.32&-27:25:49.53 & $>$18.4 & 15.12 & 11.95  &  1920 &     0.3 & \nodata &      3/0 &    60/16 & 15.35& 12.78 \\
\textrm{[BHB2007]} 9 &  17:11:21.55 & -27:27:41.70 &  12.74 &  10.57 &   8.98 &  2400 &     0.3 &  122/31 &   230/42 &   348/69 & 11.215&8.98  \\
\textrm{[BHB2007]} 12 & 17:11:25.08 & -27:24:42.67 &  16.52 &  13.61 &  11.57 &  2880 &     0.5 &  15/3   &   63/13  &    154/34 & 13.32&11.58  \\
\textrm{[BHB2007]} 13 NW&17:11:26.95 &-27:23:48.40 &  11.88 &  10.14 &   9.08 & 1440  &     0.3 &   84/32 &   126/42 &   120/3  & 10.14&9.08 \\
\textrm{[BHB2007]} 14 & 17:11:27.28 & -27:25:28.32 &  13.17 &  10.65 &  9.11  &  2400 &     0.5 &  120/21 &   351/31 &  494/111 & 10.57& 11.44 \\
\textrm{[BHB2007]} 15 & 17:11:29.43 & -27:25:36.89 &  13.28 &  11.74 &  10.70 &  2400 &     0.5 &  149/24 &   291/30 &   359/46 & 11.74&10.58  \\
\textrm{[BHB2007]} 16 & 17:11:30.35 & -27:26:28.91 &  11.91 &  10.00 &   8.89 &  1680 &     0.3 &   49/19 &    64/32 &    64/66 & 10.23&9.19  \\
\textrm{[BHB2007]} 17 & 17:11:41.00 & -27:18:36.82 &  10.64 &   8.75 &   7.54 &   960 &     0.3 &  146/18 &    150/8 &   106/19 & 8.75&7.54  \\
\textrm{[BHB2007]} 18 NE&17:11:41.73& -27:25:50.26 & \nodata& \nodata & \nodata &  480  &     0.3 &  216/37 &   287/39 &   312/50 & 10.46&9.86  \\
\textrm{[BHB2007]} 18 SW &  \nodata &      \nodata & \nodata& \nodata& \nodata&  1440 &     0.3 &   72/32 &    76/40 &    69/55 & \nodata& \nodata \\
\textrm{[BHB2007]} 19 & 17:11:43.17 & -27:30:58.54 &  14.09 &  12.75 &  11.91 &  1920 &     0.3 &   47/11 &    92/17 &   127/26 & 12.66&11.74  \\
\textrm{[BHB2007]} 20 & 17:12:00.20 & -27:20:18.10 &  10.54 &   9.64 &   9.17 &   960 &     0.3 &  119/51 &   116/51 &   146/81 & 9.641& 9.17\\
\tableline
V359 Oph & 17:08:54.28 & -27:12:33.03 & 9.73 & 9.12 & 8.80 & \nodata & \nodata & \nodata & \nodata & \nodata & \nodata &  \nodata \\
LkH$\alpha$ 345 & 17:10:48.04 & -27:40:51.19 & 9.60 & 8.69 & 8.35 & \nodata & \nodata & \nodata & \nodata & \nodata & \nodata &  \nodata \\
KK Oph A & 17:10:08.07 & -27:15:18.24 & 9.07 & 7.23 & 5.80 & \nodata & \nodata & \nodata & \nodata & \nodata & \nodata &  \nodata
\enddata
\end{deluxetable*}

%% file: t2.tex
\begin{deluxetable*}{lcccccccccc}
\tablewidth{0pt}
\tabletypesize{\tiny}
\tablecaption{IRTF/SpeX SXD Observing Log for Pipe/TW Hya Sources \label{tab:obslog_Pipe_TWHya}}
\tablehead{
  \colhead{Source} &
  \colhead{} &
  \colhead{} &
  \colhead{2MASS} &
  \colhead{2MASS} &
  \colhead{2MASS} &
  \colhead{SXD} &
  \colhead{SXD} &
  \colhead{SXD} &
  \colhead{SXD} &
  \colhead{SXD}\\
  \colhead{ID} &
  \colhead{RA} &
  \colhead{Dec} &
  \colhead{$J$} &
  \colhead{$H$} &
  \colhead{$K_{s}$} &
  \colhead{itime} &
  \colhead{slit} &
  \colhead{J S/N} &
  \colhead{H S/N} &
  \colhead{K S/N}}
\startdata
\textrm{[CLR2010]} 1  &17:14:51.08 & -27:22:39.96 &  10.26 &   8.89 &   8.16 &  360 & 0.3 &   447/55 &   654/54 &   498/84 \\
\textrm{[CLR2010]} 2 NE &  \nodata &      \nodata &\nodata & \nodata&\nodata &  480 & 0.3 &   100/44 &   107/33 &   106/36 \\
\textrm{[CLR2010]} 2 SW &17:14:56.49&-27:31:08.08 &  10.42 &   9.56 &   9.04 &  480 & 0.3 &    86/32 &   100/32 &   108/35 \\
\textrm{[CLR2010]} 3 & 17:15:59.37 & -27:00:24.75 &   6.99 &   5.72 &   5.13 &  120 & 0.3 &  507/116 &   794/19 &   935/58 \\
\textrm{[CLR2010]} 4 & 17:19:06.05 & -27:06:18.95 &  13.07 &   9.69 &   7.50 &  360 & 0.3 &     25/5 &    136/9 &   141/38 \\
\textrm{[CLR2010]} 5 & 17:20:12.47 & -26:55:09.40 &   9.12 &   7.58 &   6.67 &  720 & 0.3 &   229/30 &   241/10 &   294/24 \\
\textrm{[CLR2010]} 6 & 17:22:31.01 & -23:53:48.76 &   7.52 &   6.06 &   5.13 &  120 & 0.3 &   357/17 &    474/9 &   884/21 \\
\textrm{[CLR2010]} 7 & 17:26:15.14 & -26:44:37.31 &   9.16 &   7.13 &   5.79 &  240 & 0.3 &   180/64 &   180/20 &   182/62 \\
\textrm{[CLR2010]} 8 & 17:27:41.56 &  -26:53:46.7 &  10.12 &   7.87 &   6.19 &  240 & 0.3 &    38/15 &   146/43 &   488/49 \\
\textrm{[CLR2010]} 9 & 17:29:54.08 & -25:59:12.79 &   7.29 &   5.83 &   5.05 &  120 & 0.3 &   423/78 &   726/18 &   971/56 \\
\tableline 				   	    	      	 			    		
        TW Hya 12 &   11:21:05.6 &    -38:45:16   &   9.00 &   8.33 &   8.05 &  480 & 0.3 &  343/113 &   392/46 &   346/88 \\
     CD 29-8887 A &   11:09:14.0 &    -30:01:39   &   7.85 & \nodata&   7.18 &  360 & 0.3 &  628/110 &   776/50 &  717/100 \\
     CD 29-8887 B &   11:09:14.2 &    -30:01:38   &   9.09 &\nodata &   7.99 &  480 & 0.3 &  434/107 &   451/51 &  352/103 \\
     CD 33-7795 A &   11:31:55.4 &    -34:36:27   &   7.67 &   6.99 &   6.74 &  360 & 0.3 &  574/105 &   674/55 &   608/93 \\
     CD 34-7390 A &   11:21:17.3 &    -34:46:47   &   8.43 &   7.73 &   7.49 &  120 & 0.3 &   241/96 &   290/48 &   267/91 \\
     CD 34-7390 B &   11:21:17.5 &    -34:46:44   &\nodata &\nodata &\nodata &  240 & 0.3 &  376/116 &   473/46 &   430/87 \\
     CD 36-7429 A &   11:48:24.2 &    -37:28:49   &   8.60 &   7.95 &   7.68 &  480 & 0.3 &   392/99 &   420/47 &   343/87 \\
     CD 36-7429 B &   11:48:24.0 &    -37:28:49   &  10.06 &   9.41 &   9.14 &  480 & 0.3 &  219/115 &   265/51 &   238/86 \\
  GSC 6659-1080 A &   11:32:41.3 &    -26:51:55   &   8.34 &   7.66 &   7.43 &  120 & 0.3 &  239/104 &   285/48 &   262/81 \\
  GSC 6659-1080 B &   11:32:41.2 &    -26:52:08   &   9.84 &   9.28 &   9.01 &  240 & 0.3 &   181/77 &   202/44 &   201/49 \\
    GSC 7766-074 &   12:35:04.4 &    -41:36:39    &   9.12 &   8.48 &   8.19 &  360 & 0.3 &   227/99 &   258/51 &   224/87 \\
           GY 17 &   16:26:23.7 &  -24:43:13.9    &   9.39 &   8.40 &   7.85 &  480 & 0.3 &   371/71 &   435/57 &   377/93 \\
          GY 250 &   16:27:19.5 &  -24:41:40.5    &   9.42 &   8.63 &   8.41 &  480 & 0.3 &   540/93 &   619/48 &   514/91 \\
          GY 292 &   16:27:33.1 &  -24:41:15.7    &  11.32 &   9.13 &   7.81 &  480 & 0.3 &   136/29 &   151/41 &   114/85 \\
           GY 93 &  16:26:41.26 &    -24:40:18    &  10.77 &   9.77 &   9.27 &  480 & 0.3 &   303/55 &   330/39 &   281/42 \\
       HD 98800 A &   11:22:05.5 &  -24:46:39.7   &   6.40 &   5.76 &   5.59 &  360 & 0.3 &   863/84 &   970/46 &   783/77 \\
       Hen 3600 A &   11:10:28.0 &    -37:31:53   &   8.22 &   7.60 &   7.28 &  360 & 0.3 &  269/117 &   325/49 &   315/68 \\
       Hen 3600 B &  11:10:28.05 &  -37:31:54.2   &   8.63 &   8.07 &   7.80 &  360 & 0.3 &   213/95 &   245/49 &   223/75 \\
    SSS 1207-324 &   12:07:27.4 &    -32:47:00    &   8.62 &   8.02 &   7.75 &  360 & 0.3 &  319/112 &   364/53 &   329/78 \\
          TW Hya &   11:01:52.0 &    -34:42:17    &   8.22 &   7.56 &   7.30 &  480 & 0.3 &  501/110 &   640/57 &  615/117 \\
   TYC 7760-0283 &   12:15:30.8 &    -39:48:42    &   8.17 &   7.50 &   7.31 &  360 & 0.3 &  434/111 &   525/46 &   461/82 
\enddata
\end{deluxetable*}

%% file: t3.tex
\begin{deluxetable*}{lcccccc}
\tablewidth{0pt}
\tabletypesize{\tiny}
\tablecaption{Spectral Line Indices \label{tab:lineindices}}
\tablehead{
  \colhead{Line} &
  \colhead{Line} &
  \colhead{Line} &
  \colhead{Cont. 1} &
  \colhead{Cont. 1} &
  \colhead{Cont. 2} &
  \colhead{Cont. 2} \\
  \colhead{Name} &
  \colhead{Center ($\mu$)} &
  \colhead{Width ($\mu$)} &
  \colhead{Center ($\mu$)} &
  \colhead{Width ($\mu$)} &
  \colhead{Center ($\mu$)} &
  \colhead{Width ($\mu$)} }
\startdata
Mg 1.50 & 1.504 &  0.004 &  1.498 &  0.0045 &  1.5095 &  0.0045 \\
K 1.52 & 1.5172 &  0.004 &  1.5105 &  0.004 &  1.523 &  0.004 \\
Mg 1.58 & 1.576 &  0.004 &  1.57 &  0.004 &  1.58 &  0.004 \\
Si 1.59 &  1.59 &  0.005 &  1.586 &  0.003 &  1.594 &  0.003 \\
CO 1.62 & 1.61975 &  0.0045 &  1.615 &  0.003 &  1.628 &  0.003 \\
CO 1.66 & 1.6625 &  0.004 &  1.6565 &  0.003 &  1.67 &  0.003 \\
Al 1.67 & 1.674 &  0.006 &  1.659 &  0.003 &  1.678 &  0.002 \\
1.70 dip & 1.7075 &  0.003 &  1.704 &  0.003 &  1.7145 &  0.003 \\
Mg 1.71 & 1.7115 &  0.003 &  1.704 &  0.003 &  1.7145 &  0.003 \\
Ca 1.98 & 1.982 & 0.013 & 1.9676 & 0.005 & 1.99775 & 0.005 \\
Na 2.21 & 2.2075 & 0.007 & 2.195 & 0.004 & 2.217 & 0.006 \\
Ca 2.26 & 2.264 & 0.007 & 2.258 & 0.0045 & 2.27 & 0.0045 \\
CO 2.3 & 2.30375 & 0.0225 & 2.288 & 0.007 & 2.3185 & 0.004 \\
CO 2.34 & 2.3455 & 0.003 & 2.3425 & 0.003 & 2.349 & 0.003 \\
\enddata
\end{deluxetable*}

%% file: t4.tex
\begin{deluxetable}{lccc}
\tablewidth{0pt}
\tabletypesize{\tiny}
\tablecaption{Water Indices \label{tab:waterindices}}
\tablehead{
  \colhead{Name} &
  \colhead{Band 1} &
  \colhead{Band 2} &
  \colhead{Band 3} }
\startdata
H$_2$O-H & 1.595-1.615 & 1.68-1.70 & 1.76-1.78 \\
H$_2$O-K & 2.18-2.20 & 2.27-2.29 & 2.36-2.38 \\
\enddata
\end{deluxetable}

%% file: t5.tex
\begin{deluxetable*}{lcccccccccc}
\tablewidth{0pt}
\tabletypesize{\scriptsize}
\tablecaption{Measured Spectral  Parameters  \label{tab:B59_spts}}
\tablehead{
  \colhead{} &
  \colhead{} &
  \colhead{} &
  \colhead{} &
  \colhead{} &
  \colhead{} &
  \colhead{} &
  \colhead{} &
  \colhead{} &
  \colhead{low} &
  \colhead{} \\
  \colhead{Source} &
  \colhead{SpT}\tablenotemark{a} &
  \colhead{A$_H$} &
  \colhead{r$_J$} &
  \colhead{$\Delta$J-H} &
  \colhead{r$_H$} &
  \colhead{$\Delta$H-K} &
  \colhead{r$_K$} &
  \colhead{Br$\gamma$?} &
  \colhead{log $g$?} &
  \colhead{YSO} }
\startdata
\textrm{[BHB2007] 1} & K7  & 0.84 & 0.26 & 0.05 & 0.32 & 0.30 & 0.76 & Y & Y & Y \\
\textrm{[BHB2007] 2} & M3  & 0.68 & 0.00 & 0.00 & 0.00 & 0.00 & 0.00 & W & Y & Y \\
\textrm{[BHB2007] 3} & M6  & 1.43 & 0.00 & 0.00 & 0.00 & 0.00 & 0.00 & ? & N & ? \\
\textrm{[BHB2007] 4} & M5  & 0.74 & 0.26 & 0.10 & 0.39 & 0.16 & 0.62 & Y & Y & Y \\
\textrm{[BHB2007] 6} & M2  & 0.97 & 0.00 & 0.00 & 0.00 & 0.00 & 0.00 & N & Y & Y \\
\textrm{[BHB2007] 7} & K5:  & 2.80 & 0.45 & 0.44 & 1.16 & 0.69 & 3.11 & Y & Y & Y \\ 
\textrm{[BHB2007] 8NW}\tablenotemark{b}& K/M: & 2.41 & \nodata & \nodata & 1.20 & 1.54 & 11.5 & N & ? & ? \\ 
\textrm{[BHB2007] 8SE}\tablenotemark{b}& M: & 4.54 & \nodata & \nodata & 0.26 & 3.89 & 1.34 & N & ? & ? \\ 
\textrm{[BHB2007] 9}  & K5  & 1.34 & 0.38 & 0.39 & 0.97 & 0.61 & 2.48 & Y & Y & Y \\
\textrm{[BHB2007] 12} & M5  & 3.00 & 0.26 & 0.13 & 0.42 & 0.27 & 0.73 & Y & Y & Y \\ 
\textrm{[BHB2007] 13NW}& M2 & 2.18 & 0.15 & 0.00 & 0.15 & 0.00 & 0.15 & W? & Y & Y \\
\textrm{[BHB2007] 14} &  K5 & 2.48 & 0.38 & 0.38 & 0.96 & 0.60 & 2.43 & Y & Y & Y \\ 
\textrm{[BHB2007] 15} &  M6 & 1.74 & 0.00 & 0.01 & 0.01 & 0.01 & 0.03 & N & Y & Y \\
\textrm{[BHB2007] 16} &  K4: & 1.83 & 0.32 & 0.09 & 0.44 & 0.12 & 0.61 & N & Y & Y \\
\textrm{[BHB2007] 18NE}& M5 & 0.60 & 0.26 & 0.15 & 0.45 & 0.25 & 0.83 & Y & Y & Y \\
\textrm{[BHB2007] 18SW}& M2: & 0.76 & 0.26 & 0.12 & 0.41 & 0.27 & 0.83 & W & Y & Y \\
\textrm{[BHB2007] 19} &  M3: & 1.36 & 0.25 & 0.0 & 0.25 & 0.00 & 0.25 & Y & Y? & Y \\
\textrm{[BHB2007] 20} &  M2 & 0.48 & 0.38 & 0.04 & 0.43 & 0.13 & 0.63 & Y & Y & Y \\
\textrm{[CLR2010] 2NE} &  M3 & 0.20 & 0.25 & 0.00 & 0.25 & 0.18 & 0.48 & W? & Y? & Y? \\
\textrm{[CLR2010] 2SW} &  M5 & 0.20 & 0.25 & 0.12 & 0.40 & 0.19 & 0.67 & Y & Y? & Y? \\
\textrm{[CLR2010] 1} &  M1 & 1.30 & 0.25 & 0.00 & 0.25 & 0.00 & 0.25 & N & Y & ? \\
\enddata
\tablenotetext{a}{Spectral types flagged with colons are derived from somewhat noisy spectra, resulting in less certain derived spectral types.}
\tablenotetext{b}{Spectra are extremely noisy, and lack data in the J band; measurements of r$_J$ and $\Delta$(J-H) are impossible, and other measurements are very low quality.}
\end{deluxetable*}

%% file: t6.tex
\begin{deluxetable*}{lcccccccc}
\tablewidth{0pt}
\tabletypesize{\scriptsize}
\tablecaption{Inferred Stellar Parameters  \label{tab:inferred_params}}
\tablehead{
  \colhead{} &
  \colhead{} &
  \colhead{} &
  \colhead{} &
  \colhead{} &
  \colhead{DM98 Age} &
\colhead{DM98 Mass} &
  \colhead{B98 Age} &
\colhead{B98 Mass} \\
  \colhead{Source} &
  \colhead{T$_{eff}$} &
  \colhead{BC$_H$} &
  \colhead{Log $\frac{L_{tot}}{L_{\odot}}$} &
  \colhead{Log $\frac{L_{stel}}{L_{\odot}}$} &
  \colhead{(Myr)} &
\colhead{(M$_{\odot}$)} &
  \colhead{(Myr)} &
\colhead{(M$_{\odot}$)}
}
\startdata
\textrm{[BHB2007] 1} & 4060. & 2.11 &  0.02 & -0.10 & 0.79 & 0.49 & 4.37 & 1.14 \\
\textrm{[BHB2007] 2} & 3346. & 2.37 & -0.03 & -0.03 & $\leq$0.07 & 0.20 & $\leq$1.0 & 0.60 \\
\textrm{[BHB2007] 3} & 2841. & 2.58 & -1.38 & -1.38 & 2.59  & 0.09 & $\leq$1.0 & 0.07 \\
\textrm{[BHB2007] 4} & 3009. & 2.51 & -0.75 & -0.89 & 1.04 & 0.14 & $\leq$1.0 & 0.14 \\
\textrm{[BHB2007] 6} & 3514. & 2.27 & -0.14 & -0.14 & 0.15 & 0.24 & $\leq$1.0 &  0.62 \\
\textrm{[BHB2007] 7} & 4350. & 2.02 &  0.12 &  -0.22 & 2.71 & 0.75  & 12.02 & 1.16 \\
\textrm{[BHB2007] 9} & 4350. & 2.02 & -0.37 & -0.67 & 23.4 & 0.79 & 50.11 & 0.77 \\
\textrm{[BHB2007] 12} & 3009. & 2.51 &  -1.12 & -1.27 & 2.77 & 0.14 & 1.38 & 0.1 \\
\textrm{[BHB2007] 13NW} & 3514. & 2.27 & 0.03 & -0.03 & 0.088 & 0.23 & $\leq$1.0 & 0.68 \\
\textrm{[BHB2007] 14} & 4350. & 2.02 & 0.05 & -0.24 & 3.03 & 0.77 & 12.59 & 1.13 \\
\textrm{[BHB2007] 15} & 2841. & 2.58 &  -0.90 & -0.91 & 0.28 & 0.11 & $\leq$1.0 & 0.12 \\
\textrm{[BHB2007] 16} & 4590. & 1.98 &   0.07  & -0.09 & 2.77 & 0.88 &  13.18 & 1.24 \\
\textrm{[BHB2007] 18NE} & 3009. & 2.51 &  -0.78 & -0.94 & 1.28 & 0.14 & $\leq$1.0 & 0.13 \\
\textrm{[BHB2007] 18SW} & 3514. & 2.27 &  -0.89 & -1.04 & 6.71 &  0.34 & 13.2 & 0.46  \\
\textrm{[BHB2007] 19} & 3346. & 2.37 &  -1.37 & -1.47 & 16.6 &  0.26 &  19.1 & 0.27 \\ 
\textrm{[BHB2007] 20} & 3514. & 2.27 &  -0.44 & -0.60 &  1.37 &  0.30 &    3.47 & 0.50 \\
\textrm{V359 Oph} & 4060. & 2.11 & -0.36 & -0.36 & 2.42 & 0.61 & 10.47 & 1.02 \\
\textrm{Lk H$\alpha$ 345} & 4060. & 2.11 & -0.19 & -0.39 & 2.83 & 0.63 & 10.96 & 1.00 \\
\tableline
\textrm{[CLR2010] 2NE} & 3346. & 2.37 & -0.59 & -0.72 & 1.28 & 0.22 &   2.19 & 0.33 \\
\textrm{[CLR2010] 2SW} & 3009. & 2.51 & -0.70 & -0.84 & 0.54 & 0.13 &  $\leq$1.0  &  0.14 
\enddata
\end{deluxetable*}

%% file: ms.bbl
\begin{thebibliography}{83}
\expandafter\ifx\csname natexlab\endcsname\relax\def\natexlab#1{#1}\fi

\bibitem[{{Ali} {et~al.}(1995){Ali}, {Carr}, {Depoy}, {Frogel}, \&
  {Sellgren}}]{Ali1995}
{Ali}, B., {Carr}, J.~S., {Depoy}, D.~L., {Frogel}, J.~A., \& {Sellgren}, K.
  1995, \aj, 110, 2415

\bibitem[{{Allers} {et~al.}(2007){Allers}, {Jaffe}, {Luhman}, {Liu}, {Wilson},
  {Skrutskie}, {Nelson}, {Peterson}, {Smith}, \& {Cushing}}]{Allers2007}
{Allers}, K.~N. {et~al.} 2007, \apj, 657, 511

\bibitem[Allers et al.(2009)]{Allers2009} Allers, K.~N., et al.\ 
2009, \apj, 697, 824 

\bibitem[{{Alves} \& {Franco}(2007)}]{Alves2007a}
{Alves}, F.~O., \& {Franco}, G.~A.~P. 2007, \aap, 470, 597

\bibitem[{{Alves} {et~al.}(2008){Alves}, {Franco}, \& {Girart}}]{Alves2008}
{Alves}, F.~O., {Franco}, G.~A.~P., \& {Girart}, J.~M. 2008, \aap, 486, L13

\bibitem[{{Alves} {et~al.}(2007){Alves}, {Lombardi}, \& {Lada}}]{Alves2007}
{Alves}, J., {Lombardi}, M., \& {Lada}, C.~J. 2007, \aap, 462, L17

\bibitem[{{Aspin}(2003)}]{Aspin2003}
{Aspin}, C. 2003, \aj, 125, 1480

\bibitem[{{Ballesteros-Paredes} \& {Hartmann}(2007)}]{Ballesteros-Paredes2007}
{Ballesteros-Paredes}, J., \& {Hartmann}, L. 2007, Revista Mexicana de
  Astronomia y Astrofisica, 43, 123

\bibitem[{{Baraffe} {et~al.}(1998){Baraffe}, {Chabrier}, {Allard}, \&
  {Hauschildt}}]{Baraffe1998}
{Baraffe}, I., {Chabrier}, G., {Allard}, F., \& {Hauschildt}, P.~H. 1998, \aap,
  337, 403

\bibitem[{{Barnard} {et~al.}(1927){Barnard}, {Frost}, \&
  {Calvert}}]{Barnard1927}
{Barnard}, E.~E., {Frost}, E.~B., \& {Calvert}, M.~R. 1927, {A photographic
  atlas of selected regions of the Milky way}, ed. E.~E. {Barnard}, E.~B.
  {Frost}, \& M.~R. {Calvert}

\bibitem[{{Bastian} {et~al.}(2010){Bastian}, {Covey}, \& {Meyer}}]{Bastian2010}
{Bastian}, N., {Covey}, K.~R., \& {Meyer}, M.~R. 2010, ArXiv e-prints

\bibitem[{{Brooke} {et~al.}(2007){Brooke}, {Huard}, {Bourke}, {Boogert},
  {Allen}, {Blake}, {Evans}, {Harvey}, {Koerner}, {Mundy}, {Myers}, {Padgett},
  {Sargent}, {Stapelfeldt}, {van Dishoeck}, {Chapman}, {Cieza}, {Dunham},
  {Lai}, {Porras}, {Spiesman}, {Teuben}, {Young}, {Wahhaj}, \&
  {Lee}}]{Brooke2007}
{Brooke}, T.~Y. {et~al.} 2007, \apj, 655, 364

\bibitem[{{Carpenter}(2001)}]{Carpenter2001}
{Carpenter}, J.~M. 2001, \aj, 121, 2851

\bibitem[{Cieza {et~al.}(2005)Cieza, Kessler-Silacci, Jaffe, Harvey, \&
  II}]{Cieza2005}
Cieza, L.~A., Kessler-Silacci, J.~E., Jaffe, D.~T., Harvey, P.~M., \& II, N.
  J.~E. 2005

\bibitem[{{Covey} {et~al.}(2008){Covey}, {Hawley}, {Bochanski}, {West}, {Reid},
  {Golimowski}, {Davenport}, {Henry}, {Uomoto}, \& {Holtzman}}]{Covey2008a}
{Covey}, K.~R. {et~al.} 2008, \aj, 136, 1778

\bibitem[{{Covey} {et~al.}(2007){Covey}, {Ivezi{\'c}}, {Schlegel},
  {Finkbeiner}, {Padmanabhan}, {Lupton}, {Ag{\"u}eros}, {Bochanski}, {Hawley},
  {West}, {Seth}, {Kimball}, {Gogarten}, {Claire}, {Haggard}, {Kaib},
  {Schneider}, \& {Sesar}}]{Covey2007}
---. 2007, \aj, 134, 2398

\bibitem[{{Crutcher} {et~al.}(2009){Crutcher}, {Hakobian}, \&
  {Troland}}]{Crutcher2009}
{Crutcher}, R.~M., {Hakobian}, N., \& {Troland}, T.~H. 2009, \apj, 692, 844

\bibitem[{{Cushing} {et~al.}(2005){Cushing}, {Rayner}, \&
  {Vacca}}]{Cushing2005}
{Cushing}, M.~C., {Rayner}, J.~T., \& {Vacca}, W.~D. 2005, \apj, 623, 1115

\bibitem[{{Cushing} {et~al.}(2004){Cushing}, {Vacca}, \&
  {Rayner}}]{Cushing2004}
{Cushing}, M.~C., {Vacca}, W.~D., \& {Rayner}, J.~T. 2004, \pasp, 116, 362

\bibitem[{{D'Antona} \& {Mazzitelli}(1994)}]{Dantona1994}
{D'Antona}, F., \& {Mazzitelli}, I. 1994, \apjs, 90, 467

\bibitem[Doppmann et al.(2005)]{Doppmann2005} Doppmann, G.~W., 
Greene, T.~P., Covey, K.~R., \& Lada, C.~J.\ 2005, \aj, 130, 1145 

\bibitem[{{Elmegreen}(2000)}]{Elmegreen2000}
{Elmegreen}, B.~G. 2000, \apj, 530, 277

\bibitem[Forbrich et al.(2009)]{Forbrich2009} Forbrich, J., Lada, 
C.~J., Muench, A.~A., Alves, J., \& Lombardi, M.\ 2009, \apj, 704, 292 

\bibitem[Forbrich et al.(2010)]{Forbrich2010} Forbrich, J., Posselt, 
B., Covey, K.~R., \& Lada, C.~J.\ 2010, arXiv:1006.3556 

\bibitem[Geballe et al.(2002)]{Geballe2002} Geballe, T.~R., et al.\ 
2002, \apj, 564, 466 

\bibitem[Gorlova et al.(2003)]{Gorlova2003} Gorlova, N.~I., Meyer, 
M.~R., Rieke, G.~H., \& Liebert, J.\ 2003, \apj, 593, 1074 

\bibitem[{{Hartmann} {et~al.}(2001){Hartmann}, {Ballesteros-Paredes}, \&
  {Bergin}}]{Hartmann2001a}
{Hartmann}, L., {Ballesteros-Paredes}, J., \& {Bergin}, E.~A. 2001, \apj, 562,
  852

\bibitem[{{Herbig}(2005)}]{Herbig2005}
{Herbig}, G.~H. 2005, \aj, 130, 815

\bibitem[{{Hillenbrand}(1997)}]{Hillenbrand1997}
{Hillenbrand}, L.~A. 1997, \aj, 113, 1733

\bibitem[{{Hillenbrand} {et~al.}(2008){Hillenbrand}, {Bauermeister}, \&
  {White}}]{Hillenbrand2008}
{Hillenbrand}, L.~A., {Bauermeister}, A., \& {White}, R.~J. 2008, in
  Astronomical Society of the Pacific Conference Series, Vol. 384, 14th
  Cambridge Workshop on Cool Stars, Stellar Systems, and the Sun, ed. G.~{van
  Belle}, 200--+

\bibitem[{Indebetouw {et~al.}(2005)Indebetouw, Whitney, Johnson, \&
  Wood}]{Indebetouw2005}
Indebetouw, R., Whitney, B.~A., Johnson, K.~E., \& Wood, K. 2005

\bibitem[{{Ivanov} {et~al.}(2004){Ivanov}, {Rieke}, {Engelbracht},
  {Alonso-Herrero}, {Rieke}, \& {Luhman}}]{Ivanov2004}
{Ivanov}, V.~D., {Rieke}, M.~J., {Engelbracht}, C.~W., {Alonso-Herrero}, A.,
  {Rieke}, G.~H., \& {Luhman}, K.~L. 2004, \apjs, 151, 387

\bibitem[{{Kainulainen} {et~al.}(2009){Kainulainen}, {Lada}, {Rathborne}, \&
  {Alves}}]{Kainulainen2009}
{Kainulainen}, J., {Lada}, C.~J., {Rathborne}, J.~M., \& {Alves}, J.~F. 2009,
  \aap, 497, 399

\bibitem[{{Kenyon} {et~al.}(2008){Kenyon}, {G{\'o}mez}, \&
  {Whitney}}]{Kenyon2008}
{Kenyon}, S.~J., {G{\'o}mez}, M., \& {Whitney}, B.~A. 2008, {Low Mass Star
  Formation in the Taurus-Auriga Clouds}, ed. B.~Reipurth, 405--+

\bibitem[{{Kenyon} \& {Hartmann}(1995)}]{Kenyon1995}
{Kenyon}, S.~J., \& {Hartmann}, L. 1995, \apjs, 101, 117

\bibitem[{{Kirkpatrick} {et~al.}(1991){Kirkpatrick}, {Henry}, \&
  {McCarthy}}]{Kirkpatrick1991}
{Kirkpatrick}, J.~D., {Henry}, T.~J., \& {McCarthy}, Jr., D.~W. 1991, \apjs,
  77, 417

\bibitem[{{Kleinmann} \& {Hall}(1986)}]{Kleinmann1986}
{Kleinmann}, S.~G., \& {Hall}, D.~N.~B. 1986, \apjs, 62, 501

\bibitem[{{Klessen} {et~al.}(2000){Klessen}, {Heitsch}, \& {Mac
  Low}}]{Klessen2000}
{Klessen}, R.~S., {Heitsch}, F., \& {Mac Low}, M.-M. 2000, \apj, 535, 887

\bibitem[{{Kohoutek} \& {Wehmeyer}(2003)}]{Kohoutek2003}
{Kohoutek}, L., \& {Wehmeyer}, R. 2003, Astronomische Nachrichten, 324, 437

\bibitem[{Lada {et~al.}(2009)Lada, Lombardi, \& Alves}]{Lada2009}
Lada, C.~J., Lombardi, M., \& Alves, J.~F. 2009

\bibitem[{{Lada} {et~al.}(2008){Lada}, {Muench}, {Rathborne}, {Alves}, \&
  {Lombardi}}]{Lada2008}
{Lada}, C.~J., {Muench}, A.~A., {Rathborne}, J., {Alves}, J.~F., \& {Lombardi},
  M. 2008, \apj, 672, 410

\bibitem[{{Lan{\c c}on} \& {Wood}(2000)}]{Lancon2000}
{Lan{\c c}on}, A., \& {Wood}, P.~R. 2000, \aaps, 146, 217

\bibitem[{{Lombardi} \& {Alves}(2001)}]{Lombardi2001}
{Lombardi}, M., \& {Alves}, J. 2001, \aap, 377, 1023

\bibitem[{{Lombardi} {et~al.}(2006){Lombardi}, {Alves}, \&
  {Lada}}]{Lombardi2006}
{Lombardi}, M., {Alves}, J., \& {Lada}, C.~J. 2006, \aap, 454, 781

\bibitem[{{Luhman}(2004)}]{Luhman2004}
{Luhman}, K.~L. 2004, \apj, 617, 1216

\bibitem[{{Luhman}(2007)}]{Luhman2007}
---. 2007, \apjs, 173, 104

\bibitem[{{Luhman}(2008)}]{Luhman2008}
---. 2008, ArXiv e-prints, 808

\bibitem[{{Luhman} \& {Rieke}(1999)}]{Luhman1999}
{Luhman}, K.~L., \& {Rieke}, G.~H. 1999, \apj, 525, 440

\bibitem[{{Luhman} {et~al.}(1998){Luhman}, {Rieke}, {Lada}, \&
  {Lada}}]{Luhman1998}
{Luhman}, K.~L., {Rieke}, G.~H., {Lada}, C.~J., \& {Lada}, E.~A. 1998, \apj,
  508, 347

\bibitem[{{Luhman} {et~al.}(2006){Luhman}, {Whitney}, {Meade}, {Babler},
  {Indebetouw}, {Bracker}, \& {Churchwell}}]{Luhman2006b}
{Luhman}, K.~L., {Whitney}, B.~A., {Meade}, M.~R., {Babler}, B.~L.,
  {Indebetouw}, R., {Bracker}, S., \& {Churchwell}, E.~B. 2006, \apj, 647, 1180

\bibitem[{{Mac Low} \& {Klessen}(2004)}]{Mac-Low2004}
{Mac Low}, M.-M., \& {Klessen}, R.~S. 2004, Reviews of Modern Physics, 76, 125

\bibitem[{{Matzner}(2007)}]{Matzner2007}
{Matzner}, C.~D. 2007, \apj, 659, 1394


\bibitem[McLean et al.(2003)]{McLean2003} McLean, I.~S., McGovern, 
M.~R., Burgasser, A.~J., Kirkpatrick, J.~D., Prato, L., 
\& Kim, S.~S.\ 2003, \apj, 596, 561 

\bibitem[{{Merrill} \& {Burwell}(1950)}]{Merrill1950}
{Merrill}, P.~W., \& {Burwell}, C.~G. 1950, \apj, 112, 72

\bibitem[{{Meyer} {et~al.}(1997){Meyer}, {Calvet}, \&
  {Hillenbrand}}]{Meyer1997}
{Meyer}, M.~R., {Calvet}, N., \& {Hillenbrand}, L.~A. 1997, \aj, 114, 288

\bibitem[{{Meyer} {et~al.}(1998){Meyer}, {Edwards}, {Hinkle}, \&
  {Strom}}]{Meyer1998}
{Meyer}, M.~R., {Edwards}, S., {Hinkle}, K.~H., \& {Strom}, S.~E. 1998, \apj,
  508, 397

\bibitem[{{Morales-Calder{\'o}n} {et~al.}(2009){Morales-Calder{\'o}n},
  {Stauffer}, {Rebull}, {Whitney}, {Barrado y Navascu{\'e}s}, {Ardila}, {Song},
  {Brooke}, {Hartmann}, \& {Calvet}}]{Morales-Calderon2009}
{Morales-Calder{\'o}n}, M. {et~al.} 2009, \apj, 702, 1507

\bibitem[{{Mouschovias} \& {Ciolek}(1999)}]{Mouschovias1999}
{Mouschovias}, T.~C., \& {Ciolek}, G.~E. 1999, in NATO ASIC Proc. 540: The
  Origin of Stars and Planetary Systems, ed. C.~J. {Lada} \& N.~D. {Kylafis},
  305--+

\bibitem[{{Muench} {et~al.}(2007){Muench}, {Lada}, {Rathborne}, {Alves}, \&
  {Lombardi}}]{Muench2007}
{Muench}, A.~A., {Lada}, C.~J., {Rathborne}, J.~M., {Alves}, J.~F., \&
  {Lombardi}, M. 2007, \apj, 671, 1820

\bibitem[{{Muench} {et~al.}(2002){Muench}, {Lada}, {Lada}, \&
  {Alves}}]{Muench2002}
{Muench}, A.~A., {Lada}, E.~A., {Lada}, C.~J., \& {Alves}, J. 2002, \apj, 573,
  366

\bibitem[Mentuch et al.(2008)]{Mentuch2008} Mentuch, E., Brandeker, 
A., van Kerkwijk, M.~H., Jayawardhana, R., 
\& Hauschildt, P.~H.\ 2008, \apj, 689, 1127 

\bibitem[{{Nakamura} \& {Li}(2008)}]{Nakamura2008}
{Nakamura}, F., \& {Li}, Z.-Y. 2008, \apj, 687, 354

\bibitem[{{Nakano}(1984)}]{Nakano1984}
{Nakano}, T. 1984, Fundamentals of Cosmic Physics, 9, 139

\bibitem[{{Ochsenbein} {et~al.}(2000){Ochsenbein}, {Bauer}, \&
  {Marcout}}]{Ochsenbein2000}
{Ochsenbein}, F., {Bauer}, P., \& {Marcout}, J. 2000, \aaps, 143, 23

\bibitem[{{Offner} {et~al.}(2008){Offner}, {Klein}, \& {McKee}}]{Offner2008}
{Offner}, S.~S.~R., {Klein}, R.~I., \& {McKee}, C.~F. 2008, \apj, 686, 1174

\bibitem[{{Ojha} {et~al.}(2007){Ojha}, {Tej}, {Schultheis}, {Omont}, \&
  {Schuller}}]{Ojha2007}
{Ojha}, D.~K., {Tej}, A., {Schultheis}, M., {Omont}, A., \& {Schuller}, F.
  2007, \mnras, 381, 1219

\bibitem[{{Onishi} {et~al.}(1999){Onishi}, {Kawamura}, {Abe}, {Yamaguchi},
  {Saito}, {Moriguchi}, {Mizuno}, {Ogawa}, \& {Fukui}}]{Onishi1999}
{Onishi}, T. {et~al.} 1999, \pasj, 51, 871

\bibitem[{{Price} \& {Bate}(2009)}]{Price2009}
{Price}, D.~J., \& {Bate}, M.~R. 2009, ArXiv e-prints

\bibitem[{{Rathborne} {et~al.}(2009){Rathborne}, {Lada}, {Muench}, {Alves},
  {Kainulainen}, \& {Lombardi}}]{Rathborne2009}
{Rathborne}, J.~M., {Lada}, C.~J., {Muench}, A.~A., {Alves}, J.~F.,
  {Kainulainen}, J., \& {Lombardi}, M. 2009, \apj, 699, 742

\bibitem[{{Rathborne} {et~al.}(2008){Rathborne}, {Lada}, {Muench}, {Alves}, \&
  {Lombardi}}]{Rathborne2008}
{Rathborne}, J.~M., {Lada}, C.~J., {Muench}, A.~A., {Alves}, J.~F., \&
  {Lombardi}, M. 2008, \apjs, 174, 396

\bibitem[{{Rayner} {et~al.}(2003){Rayner}, {Toomey}, {Onaka}, {Denault},
  {Stahlberger}, {Vacca}, {Cushing}, \& {Wang}}]{Rayner2003}
{Rayner}, J.~T., {Toomey}, D.~W., {Onaka}, P.~M., {Denault}, A.~J.,
  {Stahlberger}, W.~E., {Vacca}, W.~D., {Cushing}, M.~C., \& {Wang}, S. 2003,
  \pasp, 115, 362

\bibitem[Reid et al.(2001)]{Reid2001} Reid, I.~N., Burgasser, 
A.~J., Cruz, K.~L., Kirkpatrick, J.~D., 
\& Gizis, J.~E.\ 2001, \aj, 121, 1710 

\bibitem[{{Reipurth} {et~al.}(1996){Reipurth}, {Nyman}, \&
  {Chini}}]{Reipurth1996}
{Reipurth}, B., {Nyman}, L.-A., \& {Chini}, R. 1996, \aap, 314, 258

\bibitem[{{Riaz} {et~al.}(2009){Riaz}, {Mart{\'{\i}}n}, {Bouy}, \&
  {Tata}}]{Riaz2009}
{Riaz}, B., {Mart{\'{\i}}n}, E.~L., {Bouy}, H., \& {Tata}, R. 2009, \apj, 700,
  1541

\bibitem[{{Rom{\'a}n-Z{\'u}{\~n}iga} {et~al.}(2009){Rom{\'a}n-Z{\'u}{\~n}iga},
  {Lada}, \& {Alves}}]{RomanZuniga2009}
{Rom{\'a}n-Z{\'u}{\~n}iga}, C.~G., {Lada}, C.~J., \& {Alves}, J.~F. 2009, \apj,
  704, 183

\bibitem[{{Rom{\'a}n-Z{\'u}{\~n}iga} {et~al.}(2007){Rom{\'a}n-Z{\'u}{\~n}iga},
  {Lada}, {Muench}, \& {Alves}}]{Roman-Zuniga2007}
{Rom{\'a}n-Z{\'u}{\~n}iga}, C.~G., {Lada}, C.~J., {Muench}, A., \& {Alves},
  J.~F. 2007, \apj, 664, 357

\bibitem[Sestito et 
al.(2008)]{Sestito2008} Sestito, P., Palla, F., \& Randich, S.\ 2008, \aap, 487, 965 

\bibitem[{{Shu} {et~al.}(1987){Shu}, {Adams}, \& {Lizano}}]{Shu1987}
{Shu}, F.~H., {Adams}, F.~C., \& {Lizano}, S. 1987, \araa, 25, 23

\bibitem[Slesnick et al.(2004)]{Slesnick2004} Slesnick, C.~L., 
Hillenbrand, L.~A., \& Carpenter, J.~M.\ 2004, \apj, 610, 1045 

\bibitem[{{Smith} {et~al.}(2008){Smith}, {Clark}, \& {Bonnell}}]{Smith2008}
{Smith}, R.~J., {Clark}, P.~C., \& {Bonnell}, I.~A. 2008, \mnras, 391, 1091

\bibitem[{{Smith} {et~al.}(2009){Smith}, {Clark}, \& {Bonnell}}]{Smith2009}
---. 2009, ArXiv e-prints

\bibitem[{{Stephenson} \& {Sanduleak}(1977)}]{Stephenson1977}
{Stephenson}, C.~B., \& {Sanduleak}, N. 1977, \apjs, 33, 459

\bibitem[{{Swift} \& {Welch}(2008)}]{Swift2008a}
{Swift}, J.~J., \& {Welch}, W.~J. 2008, \apjs, 174, 202

\bibitem[{{Swift} \& {Williams}(2008)}]{Swift2008}
{Swift}, J.~J., \& {Williams}, J.~P. 2008, \apj, 679, 552

\bibitem[Takagi et al.(2010)]{Takagi2010} Takagi, Y., Itoh, Y., 
\& Oasa, Y.\ 2010, \pasj, 62, 501 

\bibitem[{{Tan} {et~al.}(2006){Tan}, {Krumholz}, \& {McKee}}]{Tan2006}
{Tan}, J.~C., {Krumholz}, M.~R., \& {McKee}, C.~F. 2006, \apjl, 641, L121

\bibitem[{{The}(1964)}]{The1964}
{The}, P.-S. 1964, Contributions from the Bosscha Observervatory, 27, 1

\bibitem[{{Troland} \& {Crutcher}(2008)}]{Troland2008}
{Troland}, T.~H., \& {Crutcher}, R.~M. 2008, \apj, 680, 457

\bibitem[{{Vacca} {et~al.}(2003){Vacca}, {Cushing}, \& {Rayner}}]{Vacca2003}
{Vacca}, W.~D., {Cushing}, M.~C., \& {Rayner}, J.~T. 2003, \pasp, 115, 389

\bibitem[{{V{\'a}zquez-Semadeni} {et~al.}(2003){V{\'a}zquez-Semadeni},
  {Ballesteros-Paredes}, \& {Klessen}}]{Vazquez-Semadeni2003}
{V{\'a}zquez-Semadeni}, E., {Ballesteros-Paredes}, J., \& {Klessen}, R.~S.
  2003, \apjl, 585, L131

\bibitem[{{V{\'a}zquez-Semadeni} {et~al.}(2005){V{\'a}zquez-Semadeni}, {Kim},
  \& {Ballesteros-Paredes}}]{Vazquez-Semadeni2005a}
{V{\'a}zquez-Semadeni}, E., {Kim}, J., \& {Ballesteros-Paredes}, J. 2005,
  \apjl, 630, L49

\bibitem[{{Wallace} \& {Hinkle}(1997)}]{Wallace1997}
{Wallace}, L., \& {Hinkle}, K. 1997, \apjs, 111, 445

\bibitem[Wilking et al.(1999)]{Wilking1999} Wilking, B.~A., Greene, 
T.~P., \& Meyer, M.~R.\ 1999, \aj, 117, 469 

\bibitem[Weights et al.(2009)]{Weights2009} Weights, D.~J., Lucas, 
P.~W., Roche, P.~F., Pinfield, D.~J., 
\& Riddick, F.\ 2009, \mnras, 392, 817 

\bibitem[{{Wilking} {et~al.}(2008){Wilking}, {Gagn{\'e}}, \&
  {Allen}}]{Wilking2008}
{Wilking}, B.~A., {Gagn{\'e}}, M., \& {Allen}, L.~E. 2008, {Star Formation in
  the {$\rho$} Ophiuchi Molecular Cloud}, ed. B.~Reipurth, 351--+

\bibitem[{{Williams} {et~al.}(2009){Williams}, {Mann}, {Beaumont}, {Swift},
  {Adams}, {Hora}, {Kassis}, {Lada}, \&
  {Rom{\'a}n-Z{\'u}{\~n}iga}}]{Williams2009}
{Williams}, J.~P. {et~al.} 2009, \apj, 699, 1300

\end{thebibliography}
